\documentclass{aa}  
\usepackage{float}
\usepackage{graphicx}
\usepackage{xcolor}
\usepackage{subfigure} 
\usepackage{subcaption}
\usepackage{natbib}
                               
\usepackage{txfonts}
\usepackage{hyperref}

\begin{document}

\title{How disc initial conditions sculpt the atmospheric composition of giant planets}

\author{Angie Daniela Guzmán Franco \inst{1,2}, Sofia Savvidou\inst{2}, and Bertram Bitsch\inst{1}}

\institute{Department of Physics, University College Cork, College Rd, Cork T12 K8AF, Ireland 
\and 
Max Planck Institute for Astronomy (MPIA), Königstuhl 17, D-69117 Heidelberg
}

\date{}

\abstract{
Past studies have revealed the dependency of the disc parameters (e.g. disc mass, radius, viscosity, grain fragmentation velocity, dust-to-gas ratio) on the formation of giant planets, where more massive discs seem beneficial for giant planet formation. However, it is unclear how the different disc properties influence the composition of forming giant planets. In particular, the idea that the atmospheric abundances can trace directly the formation location of planets is put into question, due to the chemical evolution of the disc, caused by inward drifting and evaporating pebbles and chemical reactions. This complicates the idea of a simple relation between atmospheric abundances and planet formation locations. We use here planet formation simulations that include the effects of pebble drift and evaporation, and investigate how the different disc parameters influence the atmospheric composition of growing giant planets. We focus on the atmospheric C/O, C/H, O/H and S/H ratios, allowing us to probe tracers for volatiles and refractories and thus different accretion pathways of giant planets. We find that most of the disc parameters have only a limited influence on the atmospheric abundances of gas giants, except for the dust-to-gas ratio, where a larger dust-to-gas ratio results in higher atmospheric abundances. However, the atmospheric abundances are determined by the planetary formation location, even in the pebble drift and evaporation scenario. Our study suggests that volatile-rich giant exoplanets predominantly form in the inner disc regions, where they can accrete large fractions of vapour-enhanced gas. Our study shows that simulations that try to trace the origin of giant planets via their atmospheric abundances do not have to probe all disc parameters, as long as the disc parameters allow the formation of giant planets (e.g. the disc mass needs to be large enough). Consequently, our study suggests that the diversity of observed planetary compositions is a direct consequence of their formation location and migration history.
}

\authorrunning{Guzmán Franco et al.}\titlerunning{Initial conditions for atmospheric compositions}\maketitle

\section{Introduction} \label{sec:Introduction}
The recent advances in exoplanet detection have revolutionised our understanding of planetary systems, unveiling a remarkable diversity of planets that challenges and enriches existing models of planet formation. Over the past decades, the number of exoplanets has grown exponentially, reaching over 5,000 as of 2023 \citep{nasa_exoplanet_archive_2023}. These discoveries have revealed a wide range of planetary masses, sizes and orbital configurations, like hot Jupiters or super-Earths. However, the origin of these different types of planets still remains under debate \citep{Beauge2012, Izidoro2021, Turrini2021,Emsenhuber2021, bitsch2022drifting}.

Besides the planetary masses and their radii, the atmospheric composition is thought to give clues about the formation history. Originating from the work by \cite{oberg2011effects}, who describe an increase of the C/O ratio in the protoplanetary disc with increasing orbital distance - or decreasing temperature - due to the freeze out of water, CO$_2$, CH$_4$ and finally CO, many studies tried to derive the formation location of planets (e.g. \citealt{madhusudhan2017atmospheric, Pacetti_2022}) under the assumption that the disc's chemical composition does not change in time. However, simulations using the simple assumption that the disc's C/O ratio does not change in time found that the C/O ratio alone might not be enough to constrain the multi-dimensional parameter space that planet formation simulations have to cover (e.g. \citealt{Turrini2021}). Alternative constraints for planet formation simulations include atmospheric sulphur ratios (e.g. \citealt{Turrini2021, Crossfield2023}), which probe a mostly refractory component rather than carbon and oxygen, which are mostly volatile (see also \cite{schneider2021driftingpart2}). Furthermore, the bulk heavy element content of the planets determined via models that fit planetary masses and radii \citep{Thorngren_2016, Bloot2023}, can be used to inform planet formation models \citep{schneider2021drifting}.

As planets accrete material from their natal protoplanetary discs, it is essential to comprehend the chemical evolution of the disc in order to understand the composition of planets that form within it. There are two main processes that influence the chemical composition of the disc over time: 
\begin{itemize}
 \item[1)] The main carriers of volatile ices - the pebbles - drift inwards faster than the gas and can - as they evaporate in the hot inner disc regions - enrich the gas with volatile vapour, enhancing the disc's carbon and oxygen reservoirs to largely super-stellar values (e.g. \citealt{booth2017chemical, schneider2021drifting, kalyaan2023effect, Houge2025, Pacetti2025}) that can eventually be accreted by the planets \citep{schneider2021drifting, bitsch2022drifting, Penzlin2024}.

 \item[2)] As the grains move inwards, chemical reactions on the grain surfaces can alter the gas phase chemistry (e.g. \citealt{Eistrup2016}). This effect depends on the time scale differences between drift and the chemical reactions, where slower drifting grains allow for a higher efficiency of the chemical reactions \citep{Booth2019, Eistrup2022}. This directly implies that chemical reactions will become more important if grains are trapped at pressure perturbations, allowing chemical reactions to alter the disc's composition by up to an order of magnitude \citep{Pacetti2025}.
\end{itemize}

Both mechanisms can alter the disc chemical structure and thus the composition of accreting giant planets, where distinguishing between these mechanisms based on the C/O ratio alone is not possible \citep{Molliere2022, Feinstein2025}.

Planet formation in the core accretion scenario comes in two flavours, where the core either grows via planetesimal (e.g. \citealt{Pollack1996}) or pebble accretion (for a review see \citealt{johansen2017forming}). While both mechanisms can contribute to the growth of planets in the inner regions of the disc (e.g. \citealt{Izidoro2021, Batygin2023, Ogihara2024}), planetesimal accretion has difficulties explaining the growth of giant planets exterior to a few AU (e.g. \citealt{Johansen2019, Emsenhuber2021}) in contrast to pebble accretion (e.g. \citealt{Bitsch2015, Levison2015, Bitsch2023giants, Savvidou_2023}). Both approaches of giant planet formation can predict the chemical composition of giant planets \citep{cridland2016composition, madhusudhan2017atmospheric, Pekmezci2020PlanetaryRC, bitsch2022drifting, Pacetti_2022, Penzlin2024}. While the different models agree that the C/O ratio of planets should increase if their formation location is further out in the disc, they show differences in the total heavy element content of giant planets as well as in the refractory to volatile ratio in the atmosphere \citep{danti2023composition}. In particular, planets forming in discs dominated by pebbles and thus enriched by inward drifting and evaporating pebbles show a large oxygen and carbon abundance combined with an overall large heavy element content, while planets forming in discs dominated by planetesimals show a lower total heavy element enrichment. On the other hand, planets with low heavy element content formed in the planetesimal-dominated scenario should harbour a lower volatile-to-refractory ratio compared to planets with a similar total heavy element content formed in pebble-dominated scenarios. The refractory-to-volatile content could thus open a window to distinguish between the different formation scenarios \citep{danti2023composition}. \cite{Feinstein2025} gives an overview of the different assumptions related to protoplanetary discs, planet formation and how to relate those to exoplanetary atmospheres.

Here, we want to investigate how the initial disc radius, initial disc mass, $\alpha$-viscosity, starting time of embryo, dust fragmentation velocity, dust-to-gas ratio and distance from the host star influence the chemical composition of growing giant planets. \cite{Savvidou_2023} found that a large disc mass and an early formation were the most favourable conditions to form giant planets. This is caused by the fact that more massive discs that harbour larger dust masses allow a faster formation of giant planets, while planets that form too late in the disc can not grow efficiently, because the majority of the planet-forming pebbles have already drifted towards the central star. Larger viscosities lead to larger gas accretion rates as well as faster type-II migration, influencing the final mass of the giant and its final position. On the other hand, the fragmentation velocities of the dust grains - which set the pebble sizes and thus their radial drift velocities - seem to have only a small influence on the growth of the planets \citep{Savvidou_2023}. The initial birth location of the embryo also influences the growth of the planet, as pebble accretion becomes less efficient with increasing orbital distance.

We aim to explore how the C/O, C/S, and O/S ratios, along with the overall heavy element content, are shaped by these varying initial conditions and migration paths. By examining the interplay between the chemical structure of the protoplanetary disc and the dynamics of planet formation, we can gain a deeper understanding of the processes that dictate the chemical diversity of planetary systems. This knowledge is essential for interpreting observations of exoplanets and their atmospheres (e.g. \citealt{welbanks2019mass, pelletier2021water, august2023confirmation, Bardet2025, EvansSoma2025}), since it links the primordial conditions of protoplanetary discs to the final compositions of the planets that emerge from them.

We use the already existing code \texttt{chemcomp} \citep{schneider2021drifting} that contains pebble drift and evaporation at ice lines, as well as planet growth via pebble accretion coupled with migration, to study how the different disc parameters influence the growth and composition of giant planets. Our aim is to understand if the disc parameters that influence the growth of the giant planets also influence the composition of these planets, or if the composition is predominantly determined by the migration pathway of the planets rather than by the disc properties.

Our work is structured as follows: we present our modelling framework in section~\ref{sec:Methods}. We then discuss our results with respect to individual growth tracks in section~\ref{sec:tracks} before we discuss the general trends with the full planetary sample in section~\ref{sec:full}. We then discuss the implications of our findings in section~\ref{sec:Discussion}.

\section{Methods} \label{sec:Methods}

We utilised the \texttt{chemcomp} code, a semi-analytical 1D model designed for studying protoplanetary disc evolution and planetary growth, as detailed in \cite{schneider2021drifting, Schneider2024}. We will outline below the basic components of the model without going into explicit details, as these are described in \cite{schneider2021drifting}. In particular, we will analyse the planet formation simulations of \cite{Savvidou_2023} with respect to their chemical composition, indicating that the parameters are exactly the same as in \cite{Savvidou_2023}, see below. These simulations have been conducted with the \texttt{chemcomp} code version that is available on GitHub\footnote {https://github.com/AaronDavidSchneider/chemcomp}.

\subsection{Disc and dust model}

The \texttt{chemcomp} code handles the evolution of protoplanetary discs and how planets grow and migrate in them. The protoplanetary disc evolution model is governed by classic viscous evolution \citep{lynden1974evolution}, using the $\alpha$-viscosity framework \citep{shakura1973black}. The initial surface density distribution originates from the analytical solution by \cite{lynden1974evolution} and essentially scales with $\Sigma \propto r^{-1.08}$ (see section 2.3 in \cite{schneider2021drifting}). As the code contains multiple chemical species (subsection~\ref{subsec:chemistry}), the viscous evolution equation is applied to each species separately following the approach by \cite{Pavlyuchenkov2007}.

The growth of the dust grains follows the two-population approach from \cite{birnstiel2012simple}. This method distinguishes between monomer sizes and larger grain populations, with growth limited by drift, fragmentation, and drift-induced fragmentation. The inward drift velocity is then calculated by a weighted average of the velocities of the small and large grains. We also assume that the vertical stirring is governed by the vertical turbulence, which we set to $\alpha_z=10^{-4}$ \citep{pinilla2021growing}.

The mid-planet temperature is calculated via an equilibrium between viscous and stellar heating as well as radiative cooling. For simplicity, we make the assumption that the disc's temperature does not evolve in time. We also note that the grain size distribution influences the disc's cooling rate, and this in turn influences the disc properties that regulate the grain sizes in the first place (see \cite{savvidou2020influence}). We do not include this effect in detail as well, because we expect only a small impact on the disc's temperature, where changes are mostly arising from changes in the viscous heating (determined by the disc's viscosity). The exact temperature calculation of our disc model is described in Appendix B in \cite{schneider2021drifting}.

As pebbles drift inward, they reach the hotter inner disc regions and evaporate and release volatiles at their corresponding ice lines, enhancing the disc's volatile content \citep{booth2017chemical, schneider2021drifting, kalyaan2023effect}. The enhancement of the disc's volatile content is driven by the velocity differences between inward drifting grains and the viscous gas transport. As the grains drift inwards orders of magnitude faster than the gas (e.g. \citealt{Weidenschilling1977, Brauer2008}), the disc can get enhanced by orders of magnitude in volatiles, during the initial stages of the disc evolution when pebble drift is still efficient \citep{booth2017chemical, schneider2021drifting, kalyaan2023effect}. This means that the enhancement is directly regulated by the disc's viscosity and the grain sizes - set by their assumed fragmentation velocity, see \cite{bitsch2023enriching}.

In our model, we do not include the chemical evolution of gas and dust grains, because the timescales for chemical reactions on the surface of dust grains are longer than their inward drift \citep{Booth2019, Eistrup2022}. If grains are stopped at pressure perturbations, they might remain in the disc for a long enough time for grain surface chemistry reactions to change the chemical composition of the disc by up to an order of magnitude \citep{Pacetti2025}. The ALMA MAPS survey \citep{Oberg2021}, on the other hand, could not find a clear correlation between dust rings and changes in the disc chemistry. However, it is clear that including chemical reactions in planet formation simulations is one of the next necessary steps to understand the puzzle of planet formation (e.g. \citealt{Molliere2022, Pacetti2025}).

Our model also includes the re-condensation of vapour. The vapour generated by inward drifting and evaporating pebbles can also diffuse outwards and re-condense to form new pebbles. This will lead to an increase in the solid surface density at the corresponding evaporation fronts, which is proportional to the amount of material that evaporated in the first place (see section~\ref{subsec:chemistry}). The source terms for evaporation and condensation of species are integrated into the model to ensure that no more than 90\% of the local surface density is evaporated or condensed in a single time step, maintaining mass conservation throughout the simulation.

The disc dissipation is modelled by a simple exponential decay of the surface density after 3 Myrs at all orbital distances simultaneously. We use a decay time scale of 10 kyrs, so the disc's gas is emptied within a short time frame at 3 Myr (see section 2.4 in \cite{schneider2021drifting} for more details). While internal and external photoevaporation will erode the disc away in a different pattern regarding orbital distance, its influence on the disc chemical composition up until photoevaporation is negligible \citep{lienert2024changingdisccompositionsinternal, Ndugu2024}. This also applies to the composition of growing planets in discs that are subject to internal photoevaporation (Sheehan et al. in prep), motivating our simpler approach.

\subsection{Planetary evolution}

Planet formation begins with the insertion of planetary embryos into the disc at the pebble transition mass at which pebble accretion becomes efficient \citep{lambrechts2012rapid}, which corresponds to $2 \times 10^{-4}$ Earth masses at 1 AU to $10^{-1}$ Earth masses at 50 AU due to the scaling of the pebble transition mass with the disc's aspect ratio. The pebble accretion rate is calculated following the recipe of \cite{johansen2017forming}. The growth rate in itself depends not only on the mass of the planet, but also on the available amount of pebbles (the pebble surface density), leading to the fact that planetary embryos that form late in the disc with efficient pebble drift (aka smooth discs) will not grow efficiently and remain very small \citep{Savvidou_2023}, due to the massively depleted pebble mass.

As the planets grow larger, they will perturb the disc with their gravitational interactions and open small gaps in the protoplanetary disc. If the gap becomes large enough, the planet will generate a pressure bump exterior to its orbit that halts the inward flux of pebbles (e.g. \citealt{Paardekooper2006}). At this stage, the growth of the planet stops, as it has reached the so-called pebble isolation mass \citep{lambrechts2014separating, bitsch2018pebble, ataiee2018much}, which scales with $(H/r)^3$. This indicates that planetary cores in the outer disc region will need to grow more massive than planets in the inner regions to reach the pebble isolation mass ($H/r \propto r^{2/7}$). Once this mass is reached, the planet can transition into gas accretion.

We make the initial assumption that 90$\%$ of the pebble accretion rate contributes to the core growth, while $10\%$ is attributed to a primordial heavy element envelope. Once pebble isolation mass is reached, gas accretion can commence, which is modelled by the approach outlined in \cite{ndugu2021probing}. In particular, the gas accretion rate is determined by the minimum of the rates provided by \cite{ikoma2000formation} and \cite{machida2010gas}, along with the gas available from the horseshoe region around the planet \citep{ndugu2021probing}. Once the planet has opened a deep gap, its growth is limited to the disc's accretion rate $\dot{M} \propto \nu \Sigma_{\rm g}$, meaning that planets formed in discs with higher viscosities can grow to larger masses. The details of the growth model are described in section 2.6 of \cite{schneider2021drifting}.

As the planet grows, it interacts gravitationally with the protoplanetary disc and it starts to migrate (see for a recent review \citealt{Paardekooper2023}). We model the type-I migration rates of small planets via the approach of \cite{paardekooper2011torque}, which includes the effects of the Lindblad and corotation torques. In addition, we include a prescription for the dynamical torques \citep{Paardekooper2014} as well as for the heating torque \citep{BenitezLlambay2015, Jimenez2017}. As the planet starts to accrete gas, it will open deep gaps in the protoplanetary discs by gravitational effects (e.g. \citealt{Crida2007}) and by accretion \citep{BergezCasalou2020}. Once the gap is opened, the planet will start to migrate in the type-II regime, where we follow the classic viscous torques (e.g. \citealt{Baruetau2014}). Our migration prescription is described in section 2.5 of \cite{schneider2021drifting}.

\subsection{Chemistry model}
\label{subsec:chemistry}

Here we assume that the initial chemical composition of the protoplanetary disc closely resembles that of its host star (see also \cite{huhn2023accretion}), utilising solar abundances [X/H] from \cite{asplund2009chemical}. The chemical partitioning model, its molecules/solids, and their corresponding evaporation temperatures are shown in Table 2 in \cite{schneider2021drifting} and include an extension to the simpler partitioning model of \citet{bitsch2020influence} by including more elements (V, Ti, K, Na, Al, N). 

The temperature of the disc decreases with orbital distance from the central star (see Appendix B in \cite{schneider2021drifting}), which in turn affects the chemical composition of both gas and dust. When the disc temperature exceeds the condensation temperature of a given species, that species is only present in gaseous form. Conversely, when the temperature falls below this threshold, the species condenses into solid form. The transition point, where the midplane temperature equals the condensation temperature, is referred to as the evaporation front for that species \citep{schneider2021drifting}.

The outcome of our simulations is influenced by the exact chemical partitioning model, which influences the enrichment of volatiles in the disc and thus in planetary atmospheres (see \cite{schneider2021drifting}, \cite{mah2023close}). In addition, the chemical composition could change the disc structure and thus influence the growth and migration of planets (\cite{bitsch2016influence}). As we utilise here the simulations of \cite{Savvidou_2023}, where the standard chemical model of \cite{schneider2021drifting} was used, we will not investigate these points in detail. In this model, the carbon is distributed in CO, CH$_4$ and CO$_2$, which all evaporate in the outer disc regions, resulting in an increase of the C/O ratio with orbital distance.

The composition of the solid and gaseous material accreted by the planet is directly set by the composition of the material in the disc at the planet's location. Thus, as the disc composition changes (due to the inward drift of pebbles), the composition of the material that the planet accretes also changes. In particular, pebble drift and evaporation can increase the total heavy element fraction in the gas phase in the inner disc regions (where all volatiles have evaporated) to 20-30\% of the total gas content (e.g. \citealt{schneider2021drifting, bitsch2023enriching}), giving rise to very high heavy element contents of giant planets (see below). Furthermore, the planet will accrete materials of different composition once it crosses different evaporation fronts, because the chemical composition of solids and gas changes at these evaporation fronts (e.g. Fig.3 in \cite{schneider2021drifting}).

\subsection{Simulation parameters}\label{simulation_parameters}

We analyse here the simulations of \cite{Savvidou_2023}, who studied how different disc environments influence the growth of planets. In particular, \cite{Savvidou_2023} investigated how the initial disc mass, the initial disc radius, the $\alpha$-viscosity, the starting time of the embryo, the grain fragmentation velocity, the dust-to-gas ratio and the initial distance of the embryo influenced their growth. The used parameters for the individual growth tracks are listed in table ~\ref{table:standard_parameters}.

\begin{table}
\caption{\label{t7} Disc and planet parameters used in the simulations}
    \centering
    \begin{tabular}{c|c|c}
        \hline \hline
        Parameter	        & Value     & Description\\ \hline 
        $M_0$ (M$_{\odot}$) & 0.01, 0.04, 0.07, \textbf{0.1} & Initial disc mass \\
        $R_0$ (AU) & 50, 100, 150, \textbf{200} & Initial disc radius \\ 
        $\alpha$ & \textbf{0.0001}, 0.0005, 0.001 & $\alpha$-viscosity \\ 
        $t_0$ (Myr) & \textbf{0.1}, 0.5, 0.9, 1.3 & Embryo starting time \\ 
        $u_{frag}$ (ms$^{-1})$ & 1, 4, 7, \textbf{10} & Fragmentation velocity \\ 
        $f_{DG}$ & \textbf{0.015}, 0.030 & Dust-to-gas ratio \\ 
        $a$ (AU)& 1-50 \ \text{every 1 AU} & Initial orbital distance \\ \hline
    \end{tabular}
    \tablefoot{In bold, standard parameters obtained as favourable for giant planet formation from \cite{Savvidou_2023}.}
    \label{table:standard_parameters}
\end{table}

\section{Individual growth tracks} \label{sec:tracks}

We discuss in this section the individual growth tracks and how the atmospheric composition of planets changes as they grow and migrate through the disc. In particular, we investigate the [C/H], [O/H], and [S/H] ratios (normalised to the star), which give an overview of volatile and refractory accretion, considering that nearly all sulphur is in a refractory component that only evaporates in the inner disc. In addition, we show the [C/O], [C/S] and [O/S] ratios to illustrate the differences between volatile and refractory accretions as well as the total heavy element fraction of the planets. 

\cite{Savvidou_2023} varied the $\alpha$-viscosity parameters, the disc mass, the disc radius, the grain fragmentation velocity, the starting time and position of the embryo and the dust-to-gas ratio to study the growth of individual planets. \cite{Savvidou_2023} found that some parameters (e.g. disc mass, $\alpha$-viscosity, starting time of the embryo) have a large influence on whether a giant planet can grow, while other parameters only have a little influence (e.g. grain fragmentation velocity) on the growth of the planet. The presented growth tracks follow the same parameters as in \cite{Savvidou_2023}.

\subsection{Fragmentation velocity}

\begin{figure*}[htbp]
    \centering
    \includegraphics[width=0.9\textwidth]{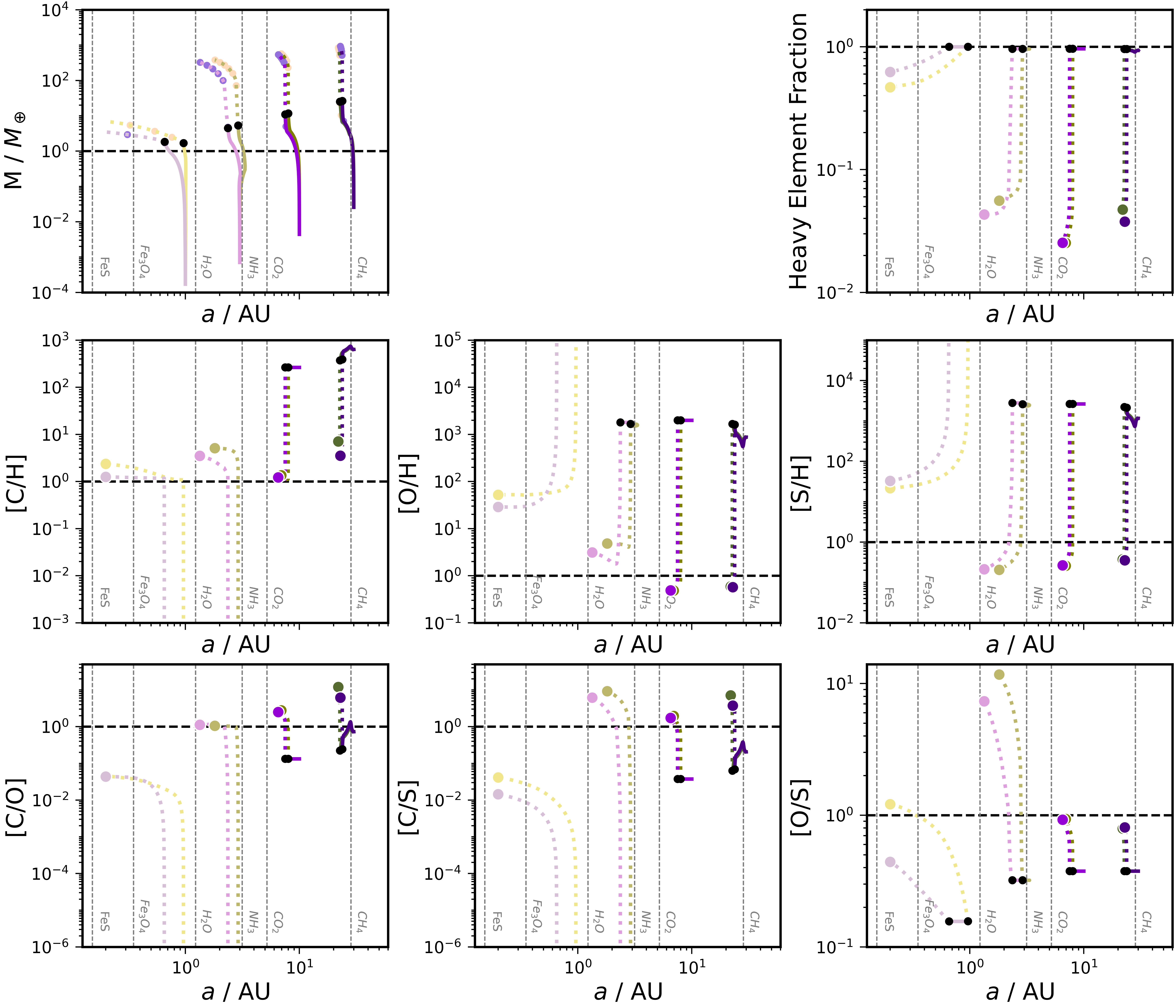}
    \caption{Evolution of planetary mass and the planetary heavy element mass fraction against position (top row) and [C/H], [O/H], [S/H], [C/O], [C/S] and [O/S] ratios (middle and bottom rows), using the standard parameters (in green) and changing to $u_{frag}$=1.0 m/s (in purple) with initial positions of 1, 3, 10 and 30 AU. The dots mark time steps of 0.5 Myr after the embryos start growing (unless the planet has reached the inner edge). The solid lines represent pebble accretion, while the dotted lines represent gas accretion, separated by a black dot.}
    \label{fig:growth_ratios_ufrag}
\end{figure*}

Fig. \ref{fig:growth_ratios_ufrag} shows the impact of a smaller fragmentation velocity ($u_{\rm frag}=1$m/s) on the formation of planets and their composition. Naively, a smaller fragmentation velocity leads to smaller grains that drift inwards slower (e.g. \citealt{Brauer2008}) and would thus enrich the disc with less evaporated volatiles \citep{bitsch2023enriching}. Consequently, a lower enrichment of planetary atmospheres could potentially be the consequence.

The smaller grains at lower fragmentation velocity result in slightly lower accretion rates (the pebble accretion rate scales with the Stokes number to the 2/3 power) with consequences for the growth and migration history of the planet (top left in Fig.~\ref{fig:growth_ratios_ufrag}). The slower growth rate results in a smaller contribution of the heating torque for the inner planets (starting at 1 and 3 AU), where the planets forming in the discs with smaller fragmentation velocity consequently migrate inwards further (by $\sim$10\%). The gas accretion rates are not affected by the smaller fragmentation velocity and thus planets starting at the same position end up with identical masses and positions that differ by up to $\sim$10\%.

The elemental abundances [C/H], [O/H], and [S/H] show initially (during the core build-up) very extreme values for all planetary starting positions independently of the grain fragmentation velocity (middle row in Fig.~\ref{fig:growth_ratios_ufrag}). This is caused by the fact that the planet does not accrete hydrogen efficiently (only via H$_2$O, H$_2$S, NH$_3$, or CH$_4$), but mostly solids containing C, O and S. The planets accrete large fractions of O and S, as these elements are always available in solid forms (e.g. FeS or MgSiO$_3$), allowing the planet to accrete these materials during core build-up. In contrast, carbon is only available in solid forms for disc temperatures below 70 K, corresponding to the CO$_2$ evaporation front. The planets forming outside the CO$_2$ evaporation front thus accrete carbon (in the form of CO$_2$) during the core build-up (which is then mixed into the envelope, see methods), while the planets forming interior to the CO$_2$ evaporation front do not accrete carbon and thus harbour initially a very low [C/H].

Similar arguments can be made about the [O/H] ratio, which is initially large for all planets, but even larger for the planet forming at $1$ AU, as it forms interior to the water ice line and is thus not accreting any hydrogen during the initial core build-up phase. Consequently, these planets (starting at 1 AU) have extremely large [O/H] ratios. The same applies to sulphur, where the outer 3 planets can accrete H$_2$S ice, while the innermost planets do not accrete any hydrogen and consequently have a very large [S/H] ratio.

As soon as the planets transition into gas accretion, the elemental abundances [C/H], [O/H], [S/H] move closer to the stellar ratios as the accretion of hydrogen commences and dilute the heavy element-enriched atmosphere. At the end of the simulations, these ratios are all similar (within 10-20\%) for the planets that become gas giants (starting at 3, 10, and 30 AU), while we observe a larger difference in the elemental ratios for the planets starting at 1 AU. The planets forming at 1 AU remain small without efficiently accreting a H/He atmosphere, and their composition is thus dominated by the core growth. The difference in atmospheric abundances reduces more for more massive planets, because the more massive planets are dominated by gas accretion that is unaffected by the grain fragmentation velocity. This trend is also reflected by the evolution of the heavy element fraction of the planet (top right panel in Fig.~\ref{fig:growth_ratios_ufrag}). While it is initially equal to $\sim$1 (by definition, the planet initially only accretes heavy elements and a tiny fraction of hydrogen hidden in e.g. H$_2$O, NH$_3$ or CH$_4$), it decreases once the planet transitions into the gas accretion phase. As the C/H, O/H, and S/H ratios of the planetary atmosphere, which essentially constitute a large fraction of the overall heavy element mass, are within 10-20\%, the heavy element fractions agree to the same level for the simulations with different fragmentation velocities.

The elemental ratios [C/O], [C/S], [O/S] show a linked behaviour (bottom row in Fig.~\ref{fig:growth_ratios_ufrag}). They remain close to unity if the corresponding X/H ratios show the same behaviour, or remain very low if one of the elements is not accreted during the core build-up. After the build-up of the core and when gas accretion starts, the ratios evolve to values closer to unity due to the accretion of evaporated volatiles and hydrogen. The ratios of [C/O], [C/S], [O/S] are very similar for the two different fragmentation velocities that were used, due to the similarity of the [C/H], [O/H] and [S/H]. We note that the [C/S] and [O/S] ratios only increase once the planet transitions into the gas accretion phase, as sulphur is mostly available in refractory form, while oxygen and carbon are accreted mostly via the gas phase due to the evaporation of the volatile carbon and oxygen carriers.

\begin{figure*}[htbp]
    \centering
    \includegraphics[width=0.9\textwidth]{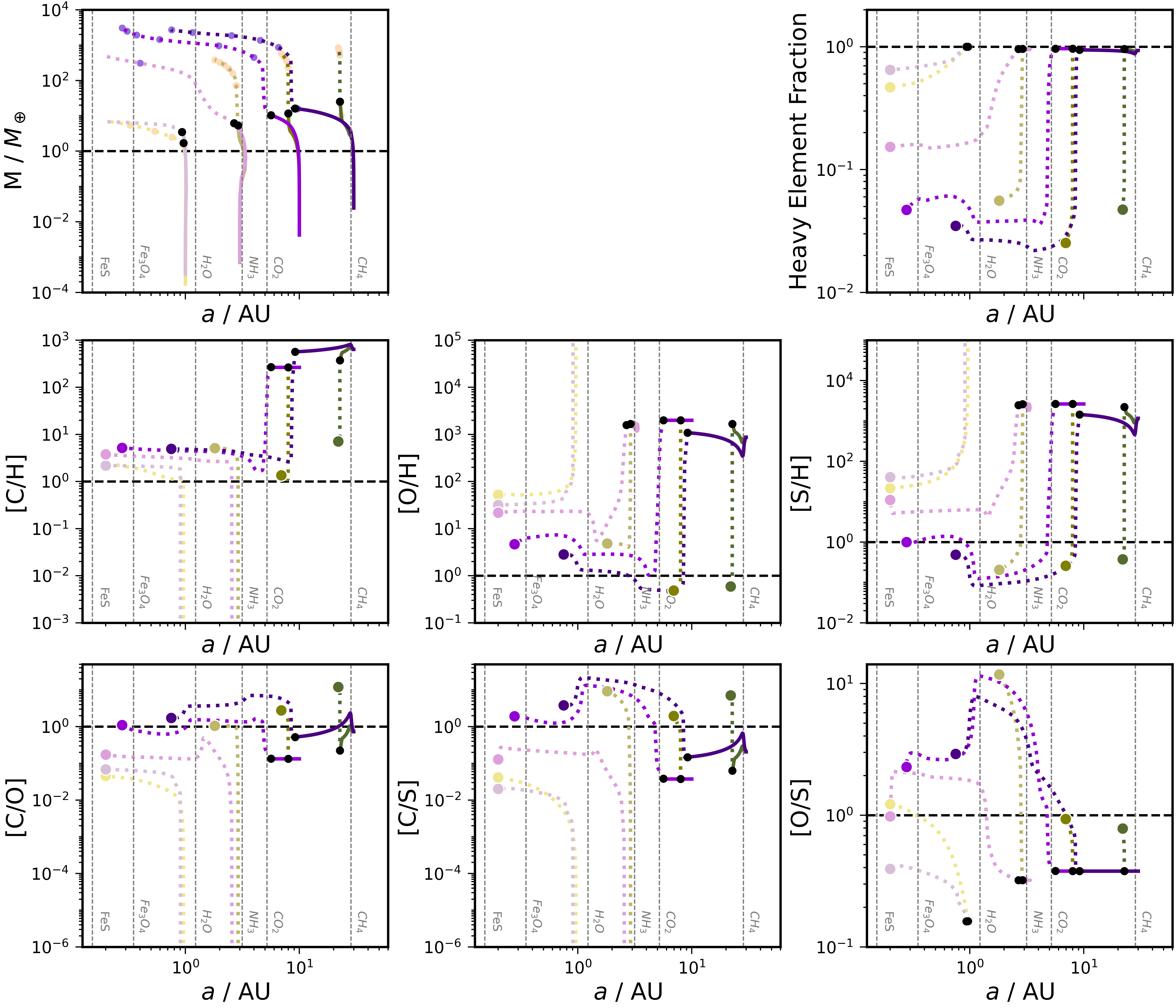}
    \caption{Same as Fig.~\ref{fig:growth_ratios_ufrag}, but now the purple colour represent planets growing in discs with $\alpha=10^{-3}$ rather than with $\alpha=10^{-4}$ (green colour).}
    \label{fig:growth_ratios_alpha}
\end{figure*}

\subsection{$\alpha$-viscosity} 

The top-left panel of Fig.~\ref{fig:growth_ratios_alpha} shows that a larger viscosity allows for a faster inward migration and a more efficient planetary growth during the gas accretion stage. The larger viscosity allows gap opening only at a later stage, thus prolonging the type-I migration phase (this can been seen during the core accretion phase for the outer planets forming at 10 and 30 AU) and at the same time, the larger viscosity will result in a faster type-II migration phase, as we model type-II migration being proportional to the disc's viscosity. In addition, the final gas accretion rates are limited by the disc's accretion rate, which is higher for higher viscosities. Consequently, the planets forming in the higher viscosity environment migrate further and grow bigger (see also \citealt{schneider2021drifting, Savvidou_2023}).

Consequently, the final abundance ratios ([C/H], [O/H], and [S/H], middle row in Fig.~\ref{fig:growth_ratios_alpha}) in the planetary atmospheres differ for planets forming at the same initial position but in discs with different viscosities. However, the initial abundance ratios during the core build-up phase agree within 10-20\%, because the cores grow to similar core masses (the pebble isolation mass is only weakly dependent on the viscosity) and the corresponding planets remain between the same evaporation fronts. The evolution of the compositions of the planetary atmospheres during core build-up follows the previous trends (Fig.~\ref{fig:growth_ratios_ufrag}).

Once the planet has reached its pebble isolation mass, it will start to accrete gas in a more efficient way. In our model, the gas accretion rate is initially set by the amount of local material within the planetary horseshoe region that is accreted, before the gas accretion rate becomes dominated by the viscous inward transport of material. A higher disc viscosity will thus allow a faster growth of the planet compared to planets forming in low viscosity environments during this accretion phase (see top left in Fig.~\ref{fig:growth_ratios_alpha}).

The atmospheric composition of the planet, on the other hand, is determined by its location in the disc, because different chemical species evaporate at different orbital distances. When comparing the planet starting at 10 AU in a disc with $\alpha=10^{-4}$ and a planet starting at 30 AU in a disc with $\alpha=10^{-3}$, our simulations show that these planets have the same atmospheric composition when compared at the same mass and same position (around 8-9 AU). This is caused by the fact that these planets accrete gaseous material with the same composition due to their formation between the same evaporation fronts. The planet forming in the disc with larger viscosity continues to grow and migrate, and eventually crosses other evaporation fronts that influence its composition by the accretion of other evaporated volatiles. Consequently, planets forming in discs with larger viscosities experience more changes in their chemical compositions due to their large-scale inward migration compared to planets forming in low viscosity environments. This suggests that migration across different evaporation fronts has a greater impact on the composition of planetary atmospheres compared to environments with different viscosities. The composition of the planets is thus determined mostly by the crossing of evaporation fronts rather than by viscosity itself. This trend is also reflected by the evolution of the heavy element fraction of the planet (top right in Fig.~\ref{fig:growth_ratios_alpha}), where planets migrating across the water ice line can receive a boost in their heavy element content due to the efficient accretion of water vapour (planets starting at 10 and 30 AU). The [C/O], [C/S], and [O/S] ratios (bottom row in Fig.~\ref{fig:growth_ratios_alpha}) reflect the trend that planets forming in discs with higher viscosity have a larger diversity in their abundance evolution as well.

\subsection{Other parameters}

The influence of other parameters, such as the initial disc mass, the initial disc radius, the starting time of the embryo, and the dust-to-gas ratio, is shown in Appendix \ref{appendix_other_parameters}, and we summarise its results here. While these parameters can strongly influence if a giant planet can grow - e.g. a late starting time will prevent the formation of giant planets \citep{Savvidou_2023} - their impact on the composition is relatively small.

As shown in \cite{Savvidou_2023}, a lower disc mass of 0.01 Solar masses does not allow the growth of giant planets. These smaller planets feature atmospheric compositions that differ by orders of magnitude compared to our standard case of higher disc mass. This is caused by the fact that the atmosphere of these small planets is dominated by the evaporation of incoming solid pebbles rather than by the accretion of volatile-rich gas (see Appendix~\ref{app:discmass}).

Delaying the starting time of the planetary embryos results in a different accretion history, where planets forming at later stages do not grow to become giant planets \citep{Savvidou_2023}, since the majority of the pebbles have already drifted inwards, resulting in a less efficient growth of the planet. We thus do not discuss the growth tracks of embryos starting to grow at 1.3 Myr, but instead show the growth tracks of embryos starting to grow at 0.9 Myr, where still a small fraction of giant planets can form (see Appendix~\ref{app:startingtime}). The corresponding atmospheric abundances can thus differ by orders of magnitude due to the different types of planets that are formed. Gas giants (above 100 Earth masses) are dominated by the accretion of vapour-enriched gas and are the common outcome of the simulations with an embryo starting time of 0.1 Myr. On the other hand, embryos starting at 0.9 Myr remain below 50 Earth masses, and their atmospheric abundances are dominated by the solid material that was accreted during the formation of the planetary core.

A smaller disc radius does not change whether giant planets can form, if they form early enough \citep{Savvidou_2023}. The corresponding atmospheric abundances are the same, as long as the planets stay within the same evaporation fronts and are compared at the same planetary mass (Appendix~\ref{app:discradius}).
    
A factor of two higher dust-to-gas ratio can lead to a more efficient planetary growth during core build-up \citep{Savvidou_2023}, see Appendix~\ref{app:dustratio}. The higher dust-to-gas ratio results in an increased amount of pebbles that drift inwards, evaporate and enrich the disc with vapour. Thus, planets forming in these discs can accrete twice as many volatiles with the gas compared to our standard simulations. However, the sulphur abundances is not influenced significantly, because sulphur is mostly accreted as solids during the build up of the core, where the higher dust-to-gas ratio does not change the total amount of solids that can be accreted (this is set by the pebble isolation mass, which is unaffected by the dust-to-gas ratio), resulting in a planetary sulphur abundance that is independent of the dust-to-gas ratio. Therefore, the [C/S] and the [O/S] ratios differ by a factor of $\sim 2$. Consequently, the individual stellar abundances are necessary as input to constrain the exact planet formation pathway, because they set the dust-to-gas ratio and the different elemental abundances, which in turn set the composition of the disc and thus of the planet (e.g. \citealt{Turrini2021, Feinstein2025}).
  
Our individual simulations already show that while different parameters can influence the growth of the giant planets, their composition is mostly affected by their migration history as long as planets are compared at the same planetary mass. This shows that the migration history has a larger influence on the planetary composition than the individual disc parameters. Consequently, planet formation simulations that want to constrain the formation history of planets do not need to probe all individual disc parameters in detail, but can rather focus on the migration history of the planet, as long as the disc conditions allow an efficient formation of the giant planets in the first place.

\section{Full planetary sample} 
\label{sec:full}

We now discuss the full planetary sample of \cite{Savvidou_2023}. By varying the individual disc (mass, radius, viscosity, grain fragmentation velocity, dust-to-gas ratio) and planet parameters (initial position and starting time), a large sample is produced. We show in Fig.~\ref{fig:mass_vs_distance_ratios} the final planetary masses as a function of the final orbital position, which we also subdivide by viscosity and planetary mass in Appendix~\ref{app:discparameters}. The different panels show the planetary C/O, C/S, and O/S ratios as well as the total heavy element mass of the planets. We first analyse the individual compositions before discussing the total heavy element content in more detail.

\subsection{Planet population}

\begin{figure*}[htbp] 
    \centering
    \begin{tabular}{cc}  
        \includegraphics[width=0.45\textwidth]{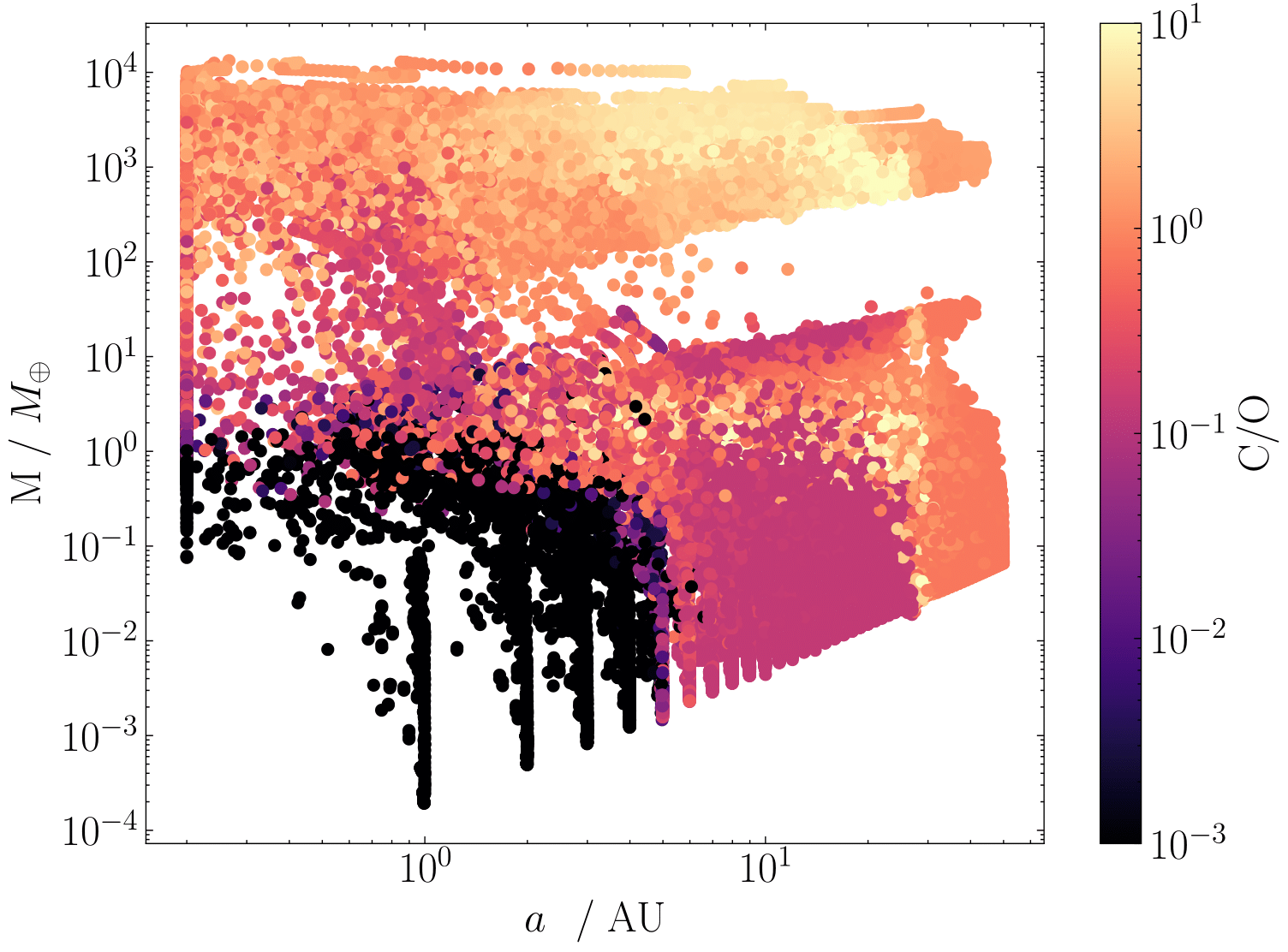} &  
        \includegraphics[width=0.45\textwidth]{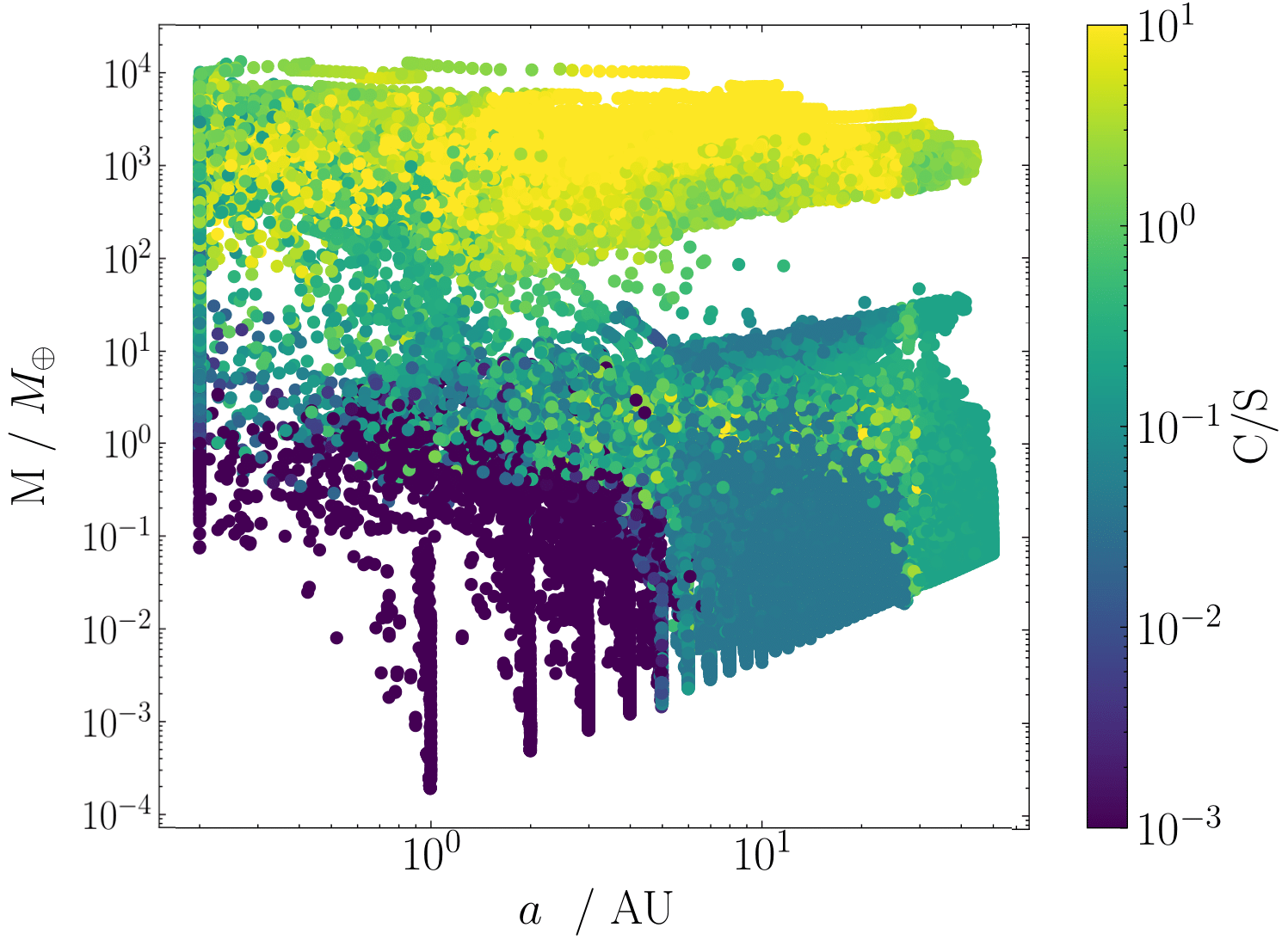} \\

        \includegraphics[width=0.45\textwidth]{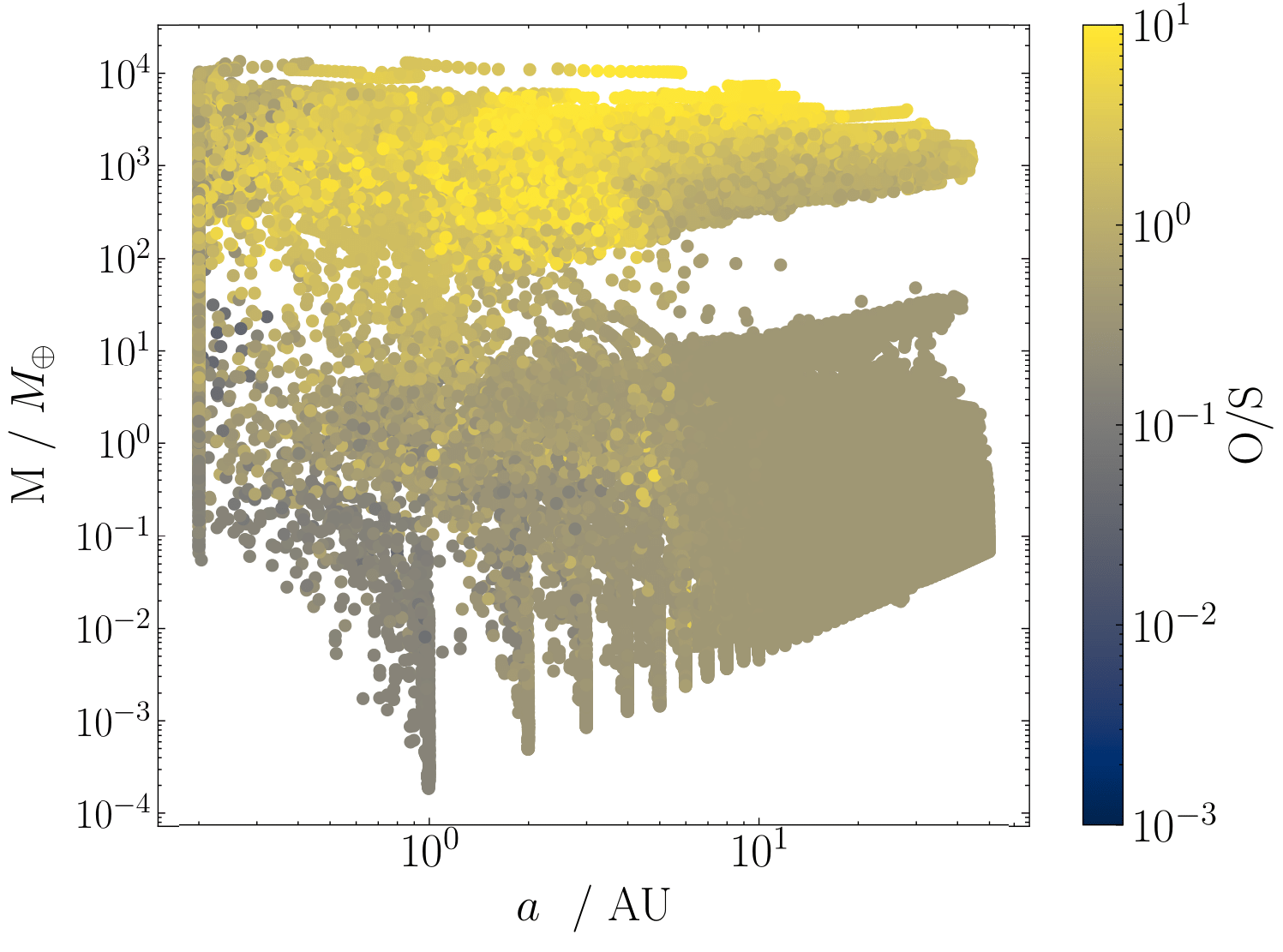} &  
        \includegraphics[width=0.45\textwidth]{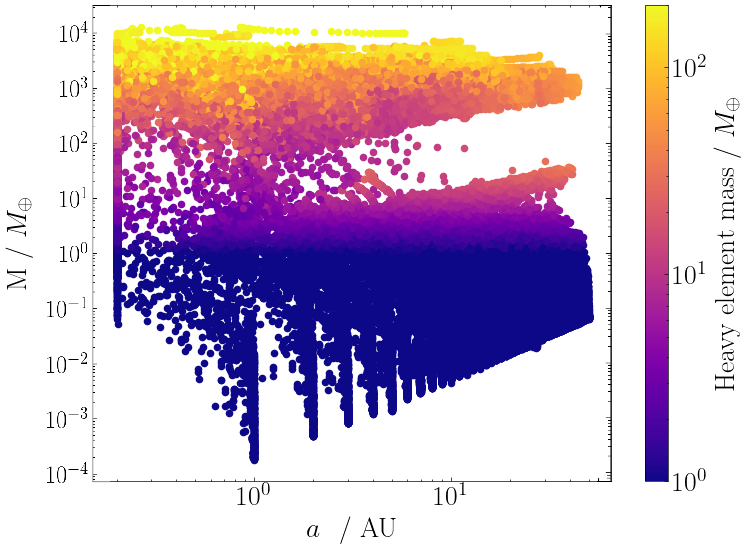} \\
 
    \end{tabular}

    \caption{Final planetary mass against final distance from the host star. The colour codes depict the C/O (top left), C/S (top right), O/S (bottom left) ratios, all normalised to the stellar value, as well as the total heavy element mass of the planets (bottom right).}
    \label{fig:mass_vs_distance_ratios}
\end{figure*}

The top left panel of Fig.~\ref{fig:mass_vs_distance_ratios} shows the atmospheric C/O ratio of the planets formed in our simulations as a function of their final mass and final orbital position. The peculiar shape of the diagram, where clear vertical lines are visible, is caused by the sampling of individual starting positions of the planetary embryos. Furthermore, a characteristic gap in planetary masses is visible between 10 and 100 Earth masses. The lower edge of this lack of planets increases with orbital distance and is caused by the core accretion paradigm. The planets first grow to pebble isolation mass (which increases with orbital distance) and then transition into runaway accretion, where the planets accrete the material in their horseshoe region efficiently and reach around 100 Earth masses. As this transition is very fast once the pebble isolation mass is reached, we expect only a small number of planets in this mass regime. This is also in agreement with other planet formation simulations (e.g. \citealt{Emsenhuber2021}).

Focusing on gas giants (planets above 100 Earth masses), the planetary C/O ratio increases with orbital distance, indicating that planets that formed further out in the disc have a larger carbon content. This is caused by two complementary effects: i) As the carbon-rich volatiles evaporate in the outer disc region, planets growing there can accrete carbon vapour for essentially their whole growth phase, and ii) Planets forming in the inner disc are dominated in their composition by evaporated water ice, which results in a low C/O ratio. The planets with final positions exterior to 30 AU harbour a C/O ratio of 1, because the only volatile that has evaporated at this distance is CO. Interior to 30 AU, the C/O ratio reaches its maximum value, because the planets have crossed interior to the CH$_4$ evaporation front, which enhances the carbon content of the planets. Moving further towards the star reduces the C/O fraction further, as only species dominated by oxygen evaporate. The here shown trend is similar to that in previous studies (e.g. \citealt{oberg2011effects, madhusudhan2017atmospheric, booth2017chemical, schneider2021drifting, Penzlin2024}, where the pebble drift and evaporation scenario allows more extreme values compared to simpler models without pebble drift and evaporation.

In the top right of Fig.~\ref{fig:mass_vs_distance_ratios} we show the C/S ratio, while the bottom left shows the O/S ratio. These ratios reveal interesting constraints about the formation history of planets, because sulphur is mostly in a refractory component (except 10\%, which is in the form of H$_2$S). This means that most of the sulphur is accreted during the formation of the planetary core rather than with the volatiles at a later stage.

The C/S and O/S ratios reveal a similar pattern compared to the C/O ratio: namely, these ratios are higher in the outer regions of the disc. The C/S ratio has its maximal value exterior to 1 AU, which is roughly the location of the H$_2$S evaporation front. In the outer disc, sulphur is only accreted during the core build-up phase, meaning it essentially remains constant during the gas accretion phase. In contrast, carbon is accreted efficiently interior to the CO$_2$ evaporation front, where CO$_2$, CH$_4$ and CO have evaporated.

Similarly, the O/S ratio is highest exterior to the water ice line due to the efficient accretion of CO$_2$. While water also contains oxygen, we actually observe a decrease of the O/S ratio interior to the water ice line, which is caused by the evaporation of H$_2$S, which takes place at the same disc temperature as water evaporation in our model.

A combination of C/O, C/S, and O/S can reveal more information about the birth locations of planets, due to the different accretion histories that are probed by O, C, and S. We will discuss the distributions in more detail below.

\subsection{Heavy element content}

The bottom right panel of Fig.~\ref{fig:mass_vs_distance_ratios} shows the total heavy element content of the planets formed in the simulations. The total heavy element content is a combination of the heavy elements in the atmosphere and in the core.  While the heavy element content is dominated by the core for smaller mass planets, it is dominated by the accretion of volatile-enriched vapour for giant planets \citep{schneider2021drifting}. As a consequence, the heavy element content of giant planets is largest if they can accrete a large fraction of their planetary envelope in the inner disc regions, where all volatile species have evaporated (e.g. \citealt{bitsch2023enriching}). Therefore, giant planets above $\sim 100$ Earth masses forming in the outer disc regions have, on average, a lower heavy element content than planets of the same mass that formed in the inner disc regions. Combining the trend of a decreasing total heavy element mass with orbital distance with the increasing C/O ratio with orbital distance (see above), implies that planets with a super-stellar C/O ratio should be less enriched in heavy elements than planets with a sub-stellar C/O ratio. This trend is as in our previous simulations \citep{schneider2021drifting} and caused by the fact that the largest heavy element content is achieved in the very inner disc regions, where water vapour is available and thus the C/O ratio is low (Fig.~\ref{fig:mass_vs_distance_ratios}).

\subsection{Atmospheric distributions}

\begin{figure*}[htbp]
    \centering
    \includegraphics[width=1.0\textwidth]{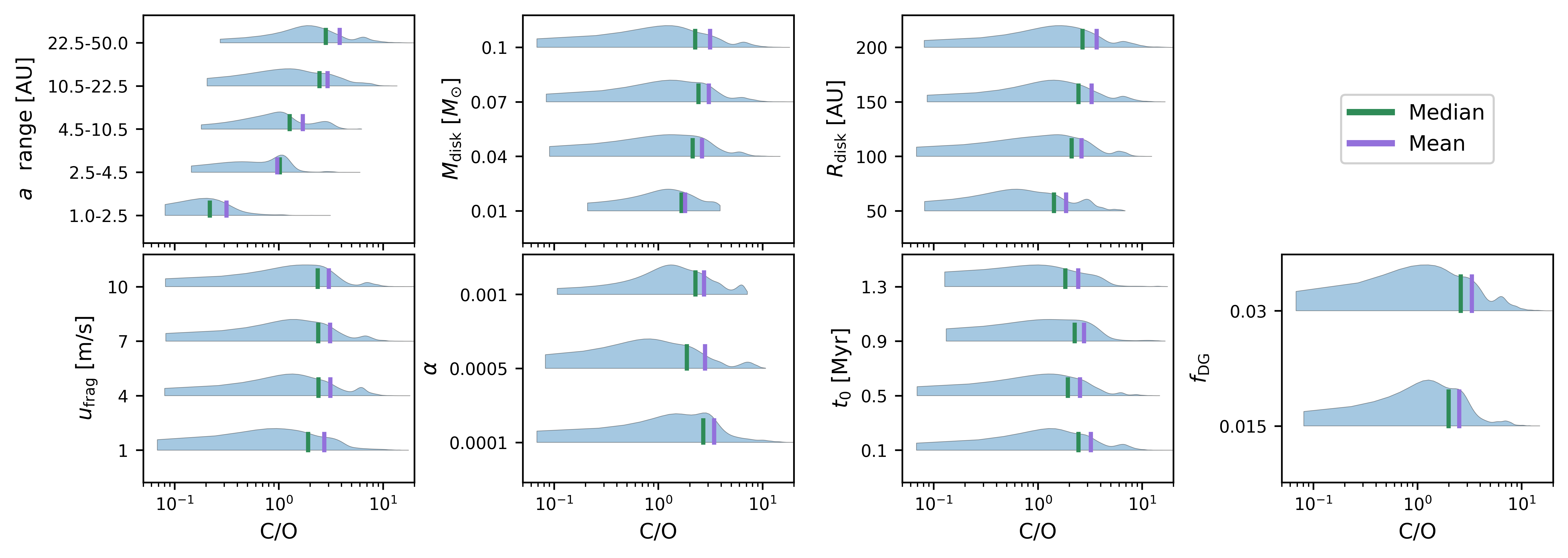}
    \caption{Violin plots of the C/O ratio ( normalised to stellar value) of the planets above 100 Earth masses within our sample. The different panels show the C/O ratio split by the different parameters that are varied in our simulations (initial orbital distance, disc mass, disc radius, fragmentation velocity, $\alpha$-viscosity, starting time and dust-to-gas ratio). The green and purple vertical lines correspond to the median and mean of the distributions.}
    \label{fig:violin_CO}
\end{figure*}

We show in Fig.~\ref{fig:violin_CO} the C/O ratio of gas giants above 100 Earth masses for our whole sample, split by the individual parameters that are varied within our simulations (table~\ref{table:standard_parameters}) while keeping the other parameters fixed. We show this data via cut violin plots, which depict the probability distributions of our sample.

In the top left panel of Fig.~\ref{fig:violin_CO}, we show the planetary C/O ratio of planets above 100 Earth masses as a function of the initial positions of the planets. The C/O ratios for the different orbital distance bins feature sub- and super-stellar ratios for all orbital distance bins. However, planets forming further out feature a higher C/O ratio (see Fig.~\ref{fig:mass_vs_distance_ratios} and Fig.~\ref{fig:discparam}), reflected by the increase of the mean and median values of the distributions with increasing planetary starting positions. This indicates that tracing the formation location of planets via their C/O ratio is more complicated by the inclusion of pebble drift and evaporation compared to simple step functions of the C/O ratio (e.g. \citealt{oberg2011effects}). We discuss the C/H, O/H and S/H distributions of our sample in Appendix~\ref{ap:ratios}.

We note that the overall shape of the distribution is influenced by our choice of the initial position of the planets. Planets forming further out in the disc feature a larger C/O ratio compared to planets forming in the inner region of the disc (see Fig.~\ref{fig:mass_vs_distance_ratios}). By choosing to simulate planets every 1 AU up to a distance of 50 AU from the central star, the distribution of the C/O ratios features only a small number of planets at sub-stellar C/O ratios, while the majority of the planets in our set of simulations have super-stellar C/O ratios. For the purpose of constraining the formation location of planets via their atmospheric abundances, we are thus mostly interested in the range of the C/O ratios for the various probed parameters rather than their absolute distributions. Our simulations suggest that the sample of planets forming within the inner 2.5 AU has a mean/median C/O distribution that is sub-stellar, indicating that planets with sub-stellar C/O ratios originate from the inner disc regions, or at least migrated into these regions.

The influence of the other parameters on the distributions of the C/O ratios reveal that these parameters (disc mass, disc radius, fragmentation velocity, $\alpha$-viscosity, starting time and dust-to-gas ratio, see Fig.~\ref{fig:violin_CO}) do not have such a significant impact on the range of the planetary C/O ratios in contrast to the formation location as the corresponding C/O ratios span the same values. Consequently, planet formation simulations that aim to extract the formation location of well-characterised giant planets do not need to probe the majority of these parameters in detail, as long as the disc properties allow the efficient formation of giant planets (e.g. the disc mass should be larger than $\sim 4\%$, \citealt{Savvidou_2023}). In the case of the low disc mass of $M_{\rm disc}=0.01M_\odot$, giant planets only form in the inner disc regions (top row in Fig.~\ref{fig:discparam}) and thus feature a lower mean C/O ratio compared to planets forming in more massive discs.

The larger dust-to-gas ratio shows a 2\textsuperscript{nd} small peak in the distribution at C/O ratios of 5-6 (bottom right in Fig.~\ref{fig:violin_CO}), in contrast to the lower dust-to-gas ratio. This is caused by the ability of lower mass discs to form planets at larger distances at high dust-to-gas ratio where the C/O ratio is larger, compared to lower dust-to-gas ratios (see \citealt{Savvidou_2023}). The C/O distributions of the other disc parameters show a similar small peak at C/O$\approx 5-6$, caused by more efficient planet formation in the outer disc regions where the planets can accrete more carbon (Fig.~\ref{fig:mass_vs_distance_ratios}). For example, the peak is only visible at early starting times, because planet formation is hindered at large distances for late starting times \citep{Savvidou_2023}. The bump is also present for the discs with larger masses, but not lower masses, as well as for discs with larger radii compared to smaller radii (e.g. if the disc only has a critical radius of 50 AU, planets have difficulties forming at large distances, as the mass reservoir at large distances is too small). The lowest grain fragmentation velocity also does not show a peak, caused by the difficulty of forming faraway planets with too small pebbles.

\section{Discussion} 
\label{sec:Discussion}

\subsection{Chemical model}

Our chemical model is a simple partitioning model, where the different elemental abundances are distributed into solids given by specific ratios \citep{schneider2021drifting}. Our model does not include chemical evolution. However, previous simulations showed that drift and evaporation should dominate over grain surface reactions, indicating that the chemical evolution should not dominate within our model (e.g. \citealt{Booth2019, Eistrup2022}). However, pebbles could be trapped by the pressure bumps generated by the planet, so that chemical reactions could change the molecular composition of the disc. A smaller grain fragmentation velocity might accelerate the chemical reactions in this case, due to the larger surface area of smaller grains (per mass). \cite{Pacetti2025} showed that the effects of chemical reactions could change the abundances of certain volatiles by up to an order of magnitude, if the grains drift inefficiently (either by being small or trapped in a pressure perturbation). Testing this is beyond the scope of this work.

Another simplicity of our chemical model is related to the carbon distribution within the disc. In our model, we do not include carbon grains and all our carbon is stored in CO, CH$_4$ and CO$_2$. In reality, there is probably a large refractory carbon grain fraction (e.g. \citealt{Lodders2003, vantHoff2020, Tabone2023, mah2023close, Pacetti2025, Houge2025}) and less methane compared to our standard assumptions. However, these changes will influence the growth of the planets in the different situations in the same way - e.g. core growth might happen faster if the carbon is situated in carbon grains compared to the very volatile methane. However, it will not influence the interplay between growth and migration of the planets. As long as the planets stay within the same evaporation fronts, independently of the exact disc parameters, their composition will be similar. The overall C/O ratio of the planets might change in a model that includes carbon grains compared to a model that does not include carbon grains, but these changes are independent of the exact disc parameters that are used.

\subsection{Planetesimal formation and accretion}

Planetesimals - km-sized objects - form if a cloud of pebbles collapses under its own gravity \citep{Youdin2005, Johansen2007, Youdin2007}. This process can happen when the local dust-to-gas ratio is enhanced to about a factor of $\sim$1 compared to the initial dust-to-gas ratio (typically of the order of 1-2\%). Different mechanisms for the formation of planetesimals have been suggested (e.g. \citealt{Drazkowska2023}), which allow the formation of planetesimals at different orbital distances and various times in protoplanetary discs. These planetesimals can then be accreted onto growing planets and change the planetary composition (e.g. \citealt{Cridland2019, Turrini2021, danti2023composition, Penzlin2024, Polychroni2025}).

Within the pebble drift and evaporation scenario, the process of planetesimal formation and accretion was studied by \citet{danti2023composition}. As planetesimals form from inward drifting pebbles, the inner disc's composition is less enhanced by vapour if planetesimals form, as they lock away the inward drifting and evaporating planets. Consequently, the overall enrichment of planetary atmospheres via vapour accretion is less if planetesimal formation is efficient \citep{danti2023composition}. 

Planetesimals will incorporate the compositions of the pebbles at the locations where they form. This directly means that planets that accrete these planetesimals will accrete compositions corresponding to the pebbles at that location. The pebbles' composition, in turn, is set by their position relative to evaporation fronts. As we have shown above, the composition of the planets accreting pebbles is determined by their location relative to evaporation fronts and not by, e.g. the disc mass. This should hold even if planetesimals are included (see also models by \citealt{danti2023composition, Penzlin2024, Polychroni2025}). On the other hand, scattering events could deliver more material to growing planetary atmospheres and thus change their composition by delivering refractory materials (e.g. \citealt{Penzlin2024, Polychroni2025}); but the range of atmospheric abundances in the case that planetesimals are included is smaller compared to the pure pebble accretion scenario, which can result in O/H and C/H ratios above 10 times stellar \citep{danti2023composition}.

\subsection{Comparison to exoplanet observations}

Our simulations show a significant trend of the atmospheric C/O ratio with initial orbital distance of the planets (Fig.~\ref{fig:violin_CO}). While relating the atmospheric composition of giant exoplanets to their formation locations requires knowledge of detailed stellar abundances for setting the initial chemical disc composition that the planet accretes from, we can compare the results of our simulations to general trends that have emerged from exoplanet atmosphere observations.

\citet{Hoch2023} showed that the atmospheric C/O ratio of transiting giant planets around solar-type stars can range from 0.2 to 2.0 normalised to the solar value by \cite{asplund2009chemical} (as our model), based on the homogeneously derived atmospheric abundances of \citet{Changeat2022}. These C/O ratios are within the ranges of our simulations, and our simulations predict that planets with super-stellar C/O ratios originate from the outer disc (Fig.~\ref{fig:mass_vs_distance_ratios}). These planets with larger C/O ratios can either migrate into the inner disc if they form in high viscosity environments (see Fig.~\ref{fig:discparam}), or could be delivered via scattering events (e.g. \citealt{Beauge2012}).

\cite{Bardet2025} re-analysed some of the spectra presented in \citet{Changeat2022} with a new model, revealing a much larger atmospheric O/H ratio for some of the hot Jupiters of the sample. They find several giant planets with atmospheric O/H ratios a few times to even 100 times the stellar value, in combination with C/O values below 0.5 stellar. Our simulations reproduce this trend for planets forming in the inner few AU, which predominantly have sub-stellar C/O ratios (Fig.~\ref{fig:violin_CO}) and super-stellar O/H (Fig.~\ref{fig:violin_H}). Our model thus predicts that these planets should have formed in the inner disc regions and probably migrated to their final position. In contrast, the observed transiting hot Jupiter planets with sub-stellar O/H and C/O$>$0.5 (corresponding to C/O$\approx$1 in our simulations that are normalised to the solar abundances with C/O$_{\rm solar} = 0.55$) would form in the outer disc and then be scattered inwards. Other planet formation models come to similar conclusions with respect to the formation location of these types of planets (e.g. \citealt{booth2017chemical, Turrini2021, Penzlin2024}).

Detailed observations of WASP-121b via the JWST revealed a heavily enriched atmosphere with O, C, and Si and a C/O ratio slightly below one \citep{EvansSoma2025} compared to its host star. Again, our model would point to a formation in the inner disc regions where planets can accrete more volatiles and also some refractories with the gas (e.g. \citealt{schneider2021driftingpart2, danti2023composition}), showing the broad consistency of our planet formation model with exoplanet observations.

\subsection{Implications for planet formation}

Constraining the formation region of giant planets is one of the top challenges in planet formation \citep{Feinstein2025}. Traditionally, it was thought that the planetary C/O ratio could help to constrain the formation region of giant planets \citep{oberg2011effects}. However, the disc's composition evolves in time via inward drifting and evaporating pebbles (e.g. \citealt{booth2017chemical, schneider2021drifting}) or also via chemical reactions (e.g. \citealt{Eistrup2016}). In addition, simulations of planet formation require initial assumptions of the formation environment (e.g. disc mass, radius) as well as of specifics related to the planetary growth (e.g. pebble or planetesimal accretion). Consequently, it is very difficult to constrain the formation of planets, as the number of assumptions might outnumber the number of available constraints. Our simulations show that a high disc mass in combination with large disc sizes and early formation times result in wide ranges of planetary C/O ratios (Fig.~\ref{fig:violin_CO}. As these conditions are beneficial for giant planet formation (e.g. \citealt{Savvidou_2023}), we expect from our model that giant planets should show a large range of C/O ratios, in agreement with observations \citep{Changeat2022, Hoch2023, Bardet2025}.

This wide range of planetary C/O ratios from within our simulations directly implies that some of the exact disc parameters (e.g. disc mass, disc radius, fragmentation velocity, viscosity, starting time) do not result in generally different outcomes, as the whole range of the planetary C/O ratios is covered within these parameters (Fig.~\ref{fig:violin_CO} and Fig.~\ref{fig:violin_H}). Consequently, as long as the disc conditions allow the formation of giant planets, one can constrain the formation location of giant planets without probing these different disc parameters (e.g. disc mass, disc radius, fragmentation velocity, viscosity, starting time) in detail. Our simulations thus show a pathway that allows for reducing the parameter space that needs to be probed to constrain the formation location of giant planets via their atmospheric abundances.

However, to constrain the formation location of giant planets it is thus necessary to use the observed composition of the planetary host star (e.g. \citealt{Cridland2019, Turrini2021}, as this will not only set the exact metallicity that can be used as input for the simulations, but will also influence the composition of the available planet forming material (e.g. \citealt{bitsch2020influence}). Furthermore, it is clear that one constraint (e.g. the C/O ratio) is not enough to constrain a multi-parameter planet formation model (e.g. \citealt{Feinstein2025}). Already using the C/H and O/H ratios in contrast to the C/O ratio allows us to constrain the formation locations of planets more precisely (see appendix~\ref{ap:ratios} and \citealt{bitsch2022drifting, danti2023composition}). We note that our analysis of the C/H, O/H and S/H in planetary atmospheres reflects the trends discussed in \cite{danti2023composition}, where planets formed in the pebble drift and evaporation scenario can reach different levels of enrichment depending on their formation location. In particular, super-solar refractory and volatile abundances are achievable in this scenario if planets migrate towards the inner disc edge (see Fig.~\ref{fig:growth_ratios_ufrag}, Fig.~\ref{fig:growth_ratios_alpha} and appendix~\ref{appendix_other_parameters}).

\section{Summary and conclusions} \label{sec:Summary_and_conclusions}

We have analysed in this paper how different disc parameters influence the composition of growing giant planets. In particular, we relied on the previous simulations of \citet{Savvidou_2023}, who investigated how the disc mass, radius, viscosity, grain fragmentation velocity, planetary starting time and dust-to-gas ratio influenced the growth efficiency of giant planets. Here, we expand on this study by investigating how these parameters influence the composition of growing giant planets. We focus on planets above 100 Earth masses, as these planets are dominated by gas rather than their cores. Smaller planets are more dominated by their core and will be investigated in upcoming works.

The wide range of planetary C/O, C/H, O/H and S/H ratios from our simulations matches the observed range of exoplanetary atmosphere abundances (e.g. \citealt{Hoch2023, Bardet2025, EvansSoma2025}. In particular, our simulations predict that hot Jupiters whose atmospheres are enhanced to super-stellar values in C, O, and S should accrete most of their material in the inner disc regions, which are enhanced in evaporated volatiles (e.g. \citealt{bitsch2023enriching}) that can be accreted with the gas. However, in line with previous studies (e.g. \citealt{Turrini2021, bitsch2022drifting}), we find that the C/O ratio alone is not a clear tracer of the planet formation location and that additional atmospheric constraints are required.

In particular, once the disc mass is large enough to form giant planets ($M_{\rm disc} >0.04M_\odot$, see \citealt{Savvidou_2023}), the disc mass has no influence on the range of planetary compositions in respect to C/O, C/H, O/H, S/H and the total heavy element content, because giant planet formation is overall efficient. The disc radius, on the other hand, has an influence mainly on the C/H ratio of the planetary atmospheres, because giant planets forming in small discs ($R_{\rm disc}=50$AU) can not form beyond $\approx$25 AU (Fig.~\ref{growth_ratios_R0}). Thus, these discs miss planets forming close to the CH$_4$ evaporation front, resulting in a lack of planets with large C/H ratios. Once the disc size is larger (e.g. $R_{\rm disc}\ge 100$ AU), there is no difference in the range of C/O, C/H, O/H, S/H and total heavy element content of the formed planets (above 100 Earth masses). The effect of the fragmentation velocity is similar in the sense that planets forming in discs with very small fragmentation velocities (1m/s) do not efficiently account for C/H ratios above $\sim 5$ (Fig.~\ref{fig:violin_H}), because the formation of wide orbit planets is hindered. On the other hand, larger grain fragmentation velocities show the same range of C/O, C/H, O/H, S/H and heavy element contents in the giant planets. The $\alpha$ viscosity parameter also has an influence on the planetary C/H ratios. Only at $\alpha=10^{-4}$ do our simulations predict sub-stellar C/H ratios. This is caused by the fact that planets forming exterior to the CO evaporation front, where carbon and oxygen are only accreted during core build-up, can remain in the outer disc due to their low inward migration velocity. The low viscosity simulations also show many planets with sub-stellar O/H, because planets forming exterior to the CO$_2$ evaporation fronts remain there and can thus only accrete the limited oxygen in the form of CO vapour during the gas accretion phase. The influence of the viscosity on the S/H and the total heavy element content, on the other hand, is negligible, as sulphur is mostly accreted with the solids. As discussed in Appendix~\ref{app:startingtime}, the starting time is mainly responsible if giant planets can form in the first place (see also \citealt{Savvidou_2023}), but the range of C/O, O/H, C/H, S/H, and heavy element content is the same as long as planets form before $t_0\le 500$ kyrs. A larger dust-to-gas ratio results in larger atmospheric abundances of C/H, O/H, and a heavy element fraction (see Appendix~\ref{app:dustratio}). However, the dust-to-gas ratio should not be treated as a free parameter in planet formation simulations, because it is set by the stellar abundances (see \citealt{Turrini2021, bitsch2022drifting, Pacetti_2022}), which also set the disc composition (e.g. \citealt{bitsch2020influence}).

We find that the composition of giant planets is mostly influenced by their starting positions and migration history. In particular, the migration of giant planets across evaporation fronts changes their composition significantly, while the individual disc parameters (e.g. disc mass, radius) only have minimal influence on the planetary composition (for planets at the same mass and same orbital distance, see Fig.~\ref{fig:discparam}). Under the assumption that our planet formation model is not missing any key ingredients, our study suggests that in order to determine the formation location of giant planets via their observed atmospheric abundances, theoretical models need at least two measured atmospheric constraints (e.g. O/H and C/H) to allow to probe the two initial variables of the starting position and the disc's viscosity, which sets the planetary migration velocity. Additionally, planet formation studies require detailed stellar abundances to eradicate the degeneracy of the dust-to-gas ratio onto planetary compositions. The combination of these constraints with state-of-the-art planet formation models can allow us to give indications of the origin of giant planets (e.g. \citealt{bitsch2022drifting}).

\begin{acknowledgements}
We thank the referee for their comments that helped to improve the clarity of our manuscript.
\end{acknowledgements}

\bibliographystyle{aa}
\bibliography{aa56632-25}

@article{Pacetti_2022,
   title={Chemical Diversity in Protoplanetary Disks and Its Impact on the Formation History of Giant Planets},
   volume={937},
   ISSN={1538-4357},
   url={http://dx.doi.org/10.3847/1538-4357/ac8b11},
   DOI={10.3847/1538-4357/ac8b11},
   number={1},
   journal={AJ},
   publisher={American Astronomical Society},
   author={Pacetti, Elenia and Turrini, Diego and Schisano, Eugenio and Molinari, Sergio and Fonte, Sergio and Politi, Romolo and Hennebelle, Patrick and Klessen, Ralf and Testi, Leonardo and Lebreuilly, Ugo},
   year={2022},
   month=sep, pages={36} }

@article{Pekmezci2020PlanetaryRC,
  title={Planetary Refractory Composition and Volatile Accretion into Gas Giants in the Protoplanetary Disks of the Sun and WASP-12},
  author={G. S. Pekmezci and Olivier Mousis and Jonathan I. Lunine},
  journal={AJ},
  year={2020},
  volume={904},
  url={https://api.semanticscholar.org/CorpusID:229415349}
}

@article{schneider2021drifting,
  title={How drifting and evaporating pebbles shape giant planets-I. Heavy element content and atmospheric C/O},
  author={Schneider, Aaron David and Bitsch, Bertram},
  journal={\aap},
  volume={654},
  pages={A71},
  year={2021},
  publisher={EDP Sciences}
}

@ARTICLE{schneider2021driftingpart2,
       author = {{Schneider}, Aaron David and {Bitsch}, Bertram},
        title = "{How drifting and evaporating pebbles shape giant planets. II. Volatiles and refractories in atmospheres}",
      journal = {\aap},
         year = 2021,
        month = oct,
       volume = {654},
          eid = {A72},
        pages = {A72},
          doi = {10.1051/0004-6361/202141096},
archivePrefix = {arXiv},
       eprint = {2109.03589},
 primaryClass = {astro-ph.EP},
       adsurl = {https://ui.adsabs.harvard.edu/abs/2021A&A...654A..72S}
}

@article{bitsch2022drifting,
  title={How drifting and evaporating pebbles shape giant planets-III. The formation of WASP-77A b and $\tau$ Bo{\"o}tis b},
  author={Bitsch, Bertram and Schneider, Aaron David and Kreidberg, Laura},
  journal={\aap},
  volume={665},
  pages={A138},
  year={2022},
  publisher={EDP Sciences}
}

@article{Savvidou_2023,
       author = {{Savvidou}, Sofia and {Bitsch}, Bertram},
        title = "{How to make giant planets via pebble accretion}",
      journal = {\aap},
         year = 2023,
        month = nov,
       volume = {679},
          eid = {A42},
        pages = {A42},
          doi = {10.1051/0004-6361/202245793},
archivePrefix = {arXiv},
       eprint = {2309.03807},
 primaryClass = {astro-ph.EP},
       adsurl = {https://ui.adsabs.harvard.edu/abs/2023A&A...679A..42S}
}

@misc{lienert2024changingdisccompositionsinternal,
      title={Changing disc compositions via internal photoevaporation}, 
      author={Julia Lena Lienert and Bertram Bitsch and Thomas Henning},
      year={2024},
      eprint={2402.09342},
      archivePrefix={arXiv},
      primaryClass={astro-ph.EP},
      url={https://arxiv.org/abs/2402.09342}, 
}

@article{Thorngren_2016,
doi = {10.3847/0004-637X/831/1/64},
url = {https://dx.doi.org/10.3847/0004-637X/831/1/64},
year = {2016},
month = {oct},
publisher = {The American Astronomical Society},
volume = {831},
number = {1},
pages = {64},
author = {Daniel P. Thorngren and Jonathan J. Fortney and Ruth A. Murray-Clay and Eric D. Lopez},
title = {THE MASS–METALLICITY RELATION FOR GIANT PLANETS},
journal = {AJ}
}

@article{lynden1974evolution,
  title={The evolution of viscous discs and the origin of the nebular variables},
  author={Lynden-Bell, Donald and Pringle, Jim E},
  journal={MNRAS},
  volume={168},
  number={3},
  pages={603--637},
  year={1974},
  publisher={The Royal Astronomical Society}
}

@article{shakura1973black,
  title={Black holes in binary systems. Observational appearance.},
  author={Shakura, Nicolai Ivanovich and Sunyaev, Rashid Alievich},
  journal={A\&A, Vol. 24, p. 337-355},
  volume={24},
  pages={337--355},
  year={1973}
}

@article{birnstiel2012simple,
  title={A simple model for the evolution of the dust population in protoplanetary disks},
  author={Birnstiel, T and Klahr, H and Ercolano, B},
  journal={\aap},
  volume={539},
  pages={A148},
  year={2012},
  publisher={EDP Sciences}
}

@article{pinilla2021growing,
  title={Growing and trapping pebbles with fragile collisions of particles in protoplanetary disks},
  author={Pinilla, Paola and Lenz, Christian T and Stammler, Sebastian M},
  journal={\aap},
  volume={645},
  pages={A70},
  year={2021},
  publisher={EDP Sciences}
}

@article{savvidou2020influence,
  title={Influence of grain growth on the thermal structure of protoplanetary discs},
  author={Savvidou, Sofia and Bitsch, Bertram and Lambrechts, Michiel},
  journal={\aap},
  volume={640},
  pages={A63},
  year={2020},
  publisher={EDP Sciences}
}

@article{lambrechts2014forming,
  title={Forming the cores of giant planets from the radial pebble flux in protoplanetary discs},
  author={Lambrechts, Michiel and Johansen, Anders},
  journal={\aap},
  volume={572},
  pages={A107},
  year={2014},
  publisher={EDP Sciences}
}

@article{lambrechts2012rapid,
  title={Rapid growth of gas-giant cores by pebble accretion},
  author={Lambrechts, Michiel and Johansen, Anders},
  journal={\aap},
  volume={544},
  pages={A32},
  year={2012},
  publisher={EDP Sciences}
}

@article{johansen2017forming,
  title={Forming planets via pebble accretion},
  author={Johansen, Anders and Lambrechts, Michiel},
  journal={Annu. Rev. Earth Planet. Sci.},
  volume={45},
  number={1},
  pages={359--387},
  year={2017},
  publisher={Annual Reviews}
}

@article{ikoma2000formation,
  title={Formation of giant planets: dependences on core accretion rate and grain opacity},
  author={Ikoma, Masahiro and Nakazawa, Kiyoshi and Emori, Hiroyuki},
  journal={AJ},
  volume={537},
  number={2},
  pages={1013},
  year={2000},
  publisher={IOP Publishing}
}

@article{machida2010gas,
  title={Gas accretion onto a protoplanet and formation of a gas giant planet},
  author={Machida, Masahiro N and Kokubo, Eiichiro and Inutsuka, Shu-ichiro and Matsumoto, Tomoaki},
  journal={MNRAS},
  volume={405},
  number={2},
  pages={1227--1243},
  year={2010},
  publisher={Blackwell Publishing Ltd Oxford, UK}
}

@article{asplund2009chemical,
  title={The chemical composition of the Sun},
  author={Asplund, Martin and Grevesse, Nicolas and Sauval, A Jacques and Scott, Pat},
  journal={ARAA},
  volume={47},
  number={1},
  pages={481--522},
  year={2009},
  publisher={Annual Reviews}
}

@article{bitsch2020influence,
  title={Influence of sub-and super-solar metallicities on the composition of solid planetary building blocks},
  author={Bitsch, Bertram and Battistini, Chiara},
  journal={\aap},
  volume={633},
  pages={A10},
  year={2020},
  publisher={EDP Sciences}
}

@misc{nasa_exoplanet_archive_2023,
  author = {{NASA Exoplanet Archive}},
  title = {NASA Exoplanet Archive},
  year = {2023},
  url = {https://exoplanetarchive.ipac.caltech.edu},
  note = {Accessed: 2024-08-27}
}

@article{madhusudhan2017atmospheric,
  title={Atmospheric signatures of giant exoplanet formation by pebble accretion},
  author={Madhusudhan, Nikku and Bitsch, Bertram and Johansen, Anders and Eriksson, Linn},
  journal={MNRAS},
  volume={469},
  number={4},
  pages={4102--4115},
  year={2017},
  publisher={Oxford University Press}
}

@article{cridland2016composition,
  title={Composition of early planetary atmospheres--I. Connecting disc astrochemistry to the formation of planetary atmospheres},
  author={Cridland, Alex J and Pudritz, Ralph E and Alessi, Matthew},
  journal={MNRAS},
  volume={461},
  number={3},
  pages={3274--3295},
  year={2016},
  publisher={Oxford University Press}
}

@article{pelletier2021water,
  title={Where is the water? Jupiter-like C/H ratio but strong H2O depletion found on $\tau$ Bo{\"o}tis b using SPIRou},
  author={Pelletier, Stefan and Benneke, Bj{\"o}rn and Darveau-Bernier, Antoine and Boucher, Anne and Cook, Neil J and Piaulet, Caroline and Coulombe, Louis-Philippe and Artigau, {\'E}tienne and Lafreni{\`e}re, David and Delisle, Simon and others},
  journal={AJ},
  volume={162},
  number={2},
  pages={73},
  year={2021},
  publisher={IOP Publishing}
}

@article{august2023confirmation,
  title={Confirmation of Subsolar Metallicity for WASP-77Ab from JWST Thermal Emission Spectroscopy},
  author={August, Prune C and Bean, Jacob L and Zhang, Michael and Lunine, Jonathan and Xue, Qiao and Line, Michael and Smith, Peter CB},
  journal={AJ Letters},
  volume={953},
  number={2},
  pages={L24},
  year={2023},
  publisher={IOP Publishing}
}

@article{welbanks2019mass,
  title={Mass--metallicity trends in transiting exoplanets from atmospheric abundances of H2O, Na, and K},
  author={Welbanks, Luis and Madhusudhan, Nikku and Allard, Nicole F and Hubeny, Ivan and Spiegelman, Fernand and Leininger, Thierry},
  journal={AJ},
  volume={887},
  number={1},
  pages={L20},
  year={2019},
  publisher={American Astronomical Society}
}

@article{oberg2011effects,
  title={The effects of snowlines on C/O in planetary atmospheres},
  author={{\"O}berg, Karin I and Murray-Clay, Ruth and Bergin, Edwin A},
  journal={AJ Letters},
  volume={743},
  number={1},
  pages={L16},
  year={2011},
  publisher={IOP Publishing}
}

@article{lambrechts2014separating,
  title={Separating gas-giant and ice-giant planets by halting pebble accretion},
  author={Lambrechts, Michiel and Johansen, Anders and Morbidelli, Alessandro},
  journal={\aap},
  volume={572},
  pages={A35},
  year={2014},
  publisher={EDP Sciences}
}

@article{ataiee2018much,
  title={How much does turbulence change the pebble isolation mass for planet formation?},
  author={Ataiee, Sareh and Baruteau, C and Alibert, Yann and Benz, Willy},
  journal={\aap},
  volume={615},
  pages={A110},
  year={2018},
  publisher={EDP Sciences}
}

@article{bitsch2018pebble,
  title={Pebble-isolation mass: Scaling law and implications for the formation of super-Earths and gas giants},
  author={Bitsch, Bertram and Morbidelli, Alessandro and Johansen, Anders and Lega, Elena and Lambrechts, Michiel and Crida, Aur{\'e}lien},
  journal={\aap},
  volume={612},
  pages={A30},
  year={2018},
  publisher={EDP Sciences}
}

@article{ndugu2021probing,
  title={Probing the impact of varied migration and gas accretion rates for the formation of giant planets in the pebble accretion scenario},
  author={Ndugu, Nelson and Bitsch, Bertram and Morbidelli, Alessandro and Crida, Aur{\'e}lien and Jurua, Edward},
  journal={MNRAS},
  volume={501},
  number={2},
  pages={2017--2028},
  year={2021},
  publisher={Oxford University Press}
}

@article{paardekooper2011torque,
  title={A torque formula for non-isothermal Type I planetary migration--II. Effects of diffusion},
  author={Paardekooper, S-J and Baruteau, C and Kley, W},
  journal={MNRAS},
  volume={410},
  number={1},
  pages={293--303},
  year={2011},
  publisher={The Royal Astronomical Society}
}

@article{huhn2023accretion,
  title={How accretion of planet-forming disks influences stellar abundances},
  author={H{\"u}hn, L-A and Bitsch, B},
  journal={\aap},
  volume={676},
  pages={A87},
  year={2023},
  publisher={EDP Sciences}
}

@article{mah2023close,
  title={Close-in ice lines and the super-stellar C/O ratio in discs around very low-mass stars},
  author={Mah, Jingyi and Bitsch, Bertram and Pascucci, Ilaria and Henning, Thomas},
  journal={\aap},
  volume={677},
  pages={L7},
  year={2023},
  publisher={EDP Sciences}
}

@article{bitsch2016influence,
  title={Influence of the water content in protoplanetary discs on planet migration and formation},
  author={Bitsch, Bertram and Johansen, Anders},
  journal={\aap},
  volume={590},
  pages={A101},
  year={2016},
  publisher={EDP Sciences}
}

@article{bitsch2023enriching,
  title={Enriching inner discs and giant planets with heavy elements},
  author={Bitsch, Bertram and Mah, Jingyi},
  journal={\aap},
  volume={679},
  pages={A11},
  year={2023},
  publisher={EDP Sciences}
}

@article{danti2023composition,
  title={Composition of giant planets: The roles of pebbles and planetesimals},
  author={Danti, Claudia and Bitsch, Bertram and Mah, Jingyi},
  journal={\aap},
  volume={679},
  pages={L7},
  year={2023},
  publisher={EDP Sciences}
}

@article{kalyaan2023effect,
  title={The Effect of Dust Evolution and Traps on Inner Disk Water Enrichment},
  author={Kalyaan, Anusha and Pinilla, Paola and Krijt, Sebastiaan and Banzatti, Andrea and Rosotti, Giovanni and Mulders, Gijs D and Lambrechts, Michiel and Long, Feng and Herczeg, Gregory J},
  journal={AJ},
  volume={954},
  number={1},
  pages={66},
  year={2023},
  publisher={IOP Publishing}
}

@article{booth2017chemical,
  title={Chemical enrichment of giant planets and discs due to pebble drift},
  author={Booth, Richard A and Clarke, Cathie J and Madhusudhan, Nikku and Ilee, John D},
  journal={MNRAS},
  volume={469},
  number={4},
  pages={3994--4011},
  year={2017},
  publisher={Oxford University Press}
}

@ARTICLE{Houge2025,
       author = {{Houge}, Adrien and {Johansen}, Anders and {Bergin}, Edwin and {Ciesla}, Fred J. and {Bitsch}, Bertram and {Lambrechts}, Michiel and {Henning}, Thomas and {Perotti}, Giulia},
        title = "{Burned to ashes: How the thermal decomposition of refractory organics in the inner protoplanetary disc impacts the gas-phase C/O ratio}",
      journal = {arXiv e-prints},
         year = 2025,
        month = may,
          eid = {arXiv:2505.20427},
        pages = {arXiv:2505.20427},
          doi = {10.48550/arXiv.2505.20427},
archivePrefix = {arXiv},
       eprint = {2505.20427},
 primaryClass = {astro-ph.EP},
       adsurl = {https://ui.adsabs.harvard.edu/abs/2025arXiv250520427H}
}

@ARTICLE{Turrini2021,
       author = {{Turrini}, D. and {Schisano}, E. and {Fonte}, S. and {Molinari}, S. and {Politi}, R. and {Fedele}, D. and {Pani{\'c}}, O. and {Kama}, M. and {Changeat}, Q. and {Tinetti}, G.},
        title = "{Tracing the Formation History of Giant Planets in Protoplanetary Disks with Carbon, Oxygen, Nitrogen, and Sulfur}",
      journal = {\apj},
         year = 2021,
        month = mar,
       volume = {909},
       number = {1},
          eid = {40},
        pages = {40},
          doi = {10.3847/1538-4357/abd6e5},
archivePrefix = {arXiv},
       eprint = {2012.14315},
 primaryClass = {astro-ph.EP},
       adsurl = {https://ui.adsabs.harvard.edu/abs/2021ApJ...909...40T}
}

@ARTICLE{Beauge2012,
       author = {{Beaug{\'e}}, C. and {Nesvorn{\'y}}, D.},
        title = "{Multiple-planet Scattering and the Origin of Hot Jupiters}",
      journal = {\apj},
         year = 2012,
        month = jun,
       volume = {751},
       number = {2},
          eid = {119},
        pages = {119},
          doi = {10.1088/0004-637X/751/2/119},
archivePrefix = {arXiv},
       eprint = {1110.4392},
 primaryClass = {astro-ph.EP},
       adsurl = {https://ui.adsabs.harvard.edu/abs/2012ApJ...751..119B}
}

@ARTICLE{Izidoro2021,
       author = {{Izidoro}, Andr{\'e} and {Bitsch}, Bertram and {Raymond}, Sean N. and {Johansen}, Anders and {Morbidelli}, Alessandro and {Lambrechts}, Michiel and {Jacobson}, Seth A.},
        title = "{Formation of planetary systems by pebble accretion and migration. Hot super-Earth systems from breaking compact resonant chains}",
      journal = {\aap},
         year = 2021,
        month = jun,
       volume = {650},
          eid = {A152},
        pages = {A152},
          doi = {10.1051/0004-6361/201935336},
archivePrefix = {arXiv},
       eprint = {1902.08772},
 primaryClass = {astro-ph.EP},
       adsurl = {https://ui.adsabs.harvard.edu/abs/2021A&A...650A.152I}
}

@ARTICLE{Emsenhuber2021,
       author = {{Emsenhuber}, Alexandre and {Mordasini}, Christoph and {Burn}, Remo and {Alibert}, Yann and {Benz}, Willy and {Asphaug}, Erik},
        title = "{The New Generation Planetary Population Synthesis (NGPPS). I. Bern global model of planet formation and evolution, model tests, and emerging planetary systems}",
      journal = {\aap},
         year = 2021,
        month = dec,
       volume = {656},
          eid = {A69},
        pages = {A69},
          doi = {10.1051/0004-6361/202038553},
archivePrefix = {arXiv},
       eprint = {2007.05561},
 primaryClass = {astro-ph.EP},
       adsurl = {https://ui.adsabs.harvard.edu/abs/2021A&A...656A..69E}
}

@ARTICLE{Crossfield2023,
       author = {{Crossfield}, Ian J.~M.},
        title = "{Volatile-to-sulfur Ratios Can Recover a Gas Giant's Accretion History}",
      journal = {\apjl},
         year = 2023,
        month = jul,
       volume = {952},
       number = {1},
          eid = {L18},
        pages = {L18},
          doi = {10.3847/2041-8213/ace35f},
archivePrefix = {arXiv},
       eprint = {2303.17622},
 primaryClass = {astro-ph.EP},
       adsurl = {https://ui.adsabs.harvard.edu/abs/2023ApJ...952L..18C}
}

@ARTICLE{Bloot2023,
       author = {{Bloot}, S. and {Miguel}, Y. and {Bazot}, M. and {Howard}, S.},
        title = "{Exoplanet interior retrievals: core masses and metallicities from atmospheric abundances}",
      journal = {\mnras},
         year = 2023,
        month = aug,
       volume = {523},
       number = {4},
        pages = {6282-6292},
          doi = {10.1093/mnras/stad1873},
archivePrefix = {arXiv},
       eprint = {2306.11354},
 primaryClass = {astro-ph.EP},
       adsurl = {https://ui.adsabs.harvard.edu/abs/2023MNRAS.523.6282B}
}

@ARTICLE{Penzlin2024,
       author = {{Penzlin}, Anna B.~T. and {Booth}, Richard A. and {Kirk}, James and {Owen}, James E. and {Ahrer}, E. and {Christie}, Duncan A. and {Claringbold}, Alastair B. and {Esparza-Borges}, Emma and {L{\'o}pez-Morales}, M. and {Mayne}, N.~J. and {McCormack}, Mason and {Meech}, Annabella and {Panwar}, Vatsal and {Powell}, Diana and {Sergeev}, Denis E. and {Taylor}, Jake and {Wheatley}, Peter J. and {Zamyatina}, Maria},
        title = "{BOWIE-ALIGN: how formation and migration histories of giant planets impact atmospheric compositions}",
      journal = {\mnras},
         year = 2024,
        month = nov,
       volume = {535},
       number = {1},
        pages = {171-186},
          doi = {10.1093/mnras/stae2362},
archivePrefix = {arXiv},
       eprint = {2407.03199},
 primaryClass = {astro-ph.EP},
       adsurl = {https://ui.adsabs.harvard.edu/abs/2024MNRAS.535..171P}
}

@ARTICLE{Eistrup2016,
       author = {{Eistrup}, Christian and {Walsh}, Catherine and {van Dishoeck}, Ewine F.},
        title = "{Setting the volatile composition of (exo)planet-building material. Does chemical evolution in disk midplanes matter?}",
      journal = {\aap},
         year = 2016,
        month = nov,
       volume = {595},
          eid = {A83},
        pages = {A83},
          doi = {10.1051/0004-6361/201628509},
archivePrefix = {arXiv},
       eprint = {1607.06710},
 primaryClass = {astro-ph.EP},
       adsurl = {https://ui.adsabs.harvard.edu/abs/2016A&A...595A..83E}
}

@ARTICLE{Eistrup2022,
       author = {{Eistrup}, Christian and {Henning}, Thomas},
        title = "{Chemical evolution in ices on drifting, planet-forming pebbles}",
      journal = {\aap},
         year = 2022,
        month = nov,
       volume = {667},
          eid = {A160},
        pages = {A160},
          doi = {10.1051/0004-6361/202243982},
archivePrefix = {arXiv},
       eprint = {2208.07390},
 primaryClass = {astro-ph.EP},
       adsurl = {https://ui.adsabs.harvard.edu/abs/2022A&A...667A.160E}
}

@ARTICLE{Booth2019,
       author = {{Booth}, R.~A. and {Ilee}, J.~D.},
        title = "{Planet-forming material in a protoplanetary disc: the interplay between chemical evolution and pebble drift}",
      journal = {\mnras},
         year = 2019,
        month = aug,
       volume = {487},
       number = {3},
        pages = {3998-4011},
          doi = {10.1093/mnras/stz1488},
archivePrefix = {arXiv},
       eprint = {1905.12639},
 primaryClass = {astro-ph.EP},
       adsurl = {https://ui.adsabs.harvard.edu/abs/2019MNRAS.487.3998B}
}

@ARTICLE{Molliere2022,
       author = {{Molli{\`e}re}, Paul and {Molyarova}, Tamara and {Bitsch}, Bertram and {Henning}, Thomas and {Schneider}, Aaron and {Kreidberg}, Laura and {Eistrup}, Christian and {Burn}, Remo and {Nasedkin}, Evert and {Semenov}, Dmitry and {Mordasini}, Christoph and {Schlecker}, Martin and {Schwarz}, Kamber R. and {Lacour}, Sylvestre and {Nowak}, Mathias and {Schulik}, Matth{\"a}us},
        title = "{Interpreting the Atmospheric Composition of Exoplanets: Sensitivity to Planet Formation Assumptions}",
      journal = {\apj},
         year = 2022,
        month = jul,
       volume = {934},
       number = {1},
          eid = {74},
        pages = {74},
          doi = {10.3847/1538-4357/ac6a56},
archivePrefix = {arXiv},
       eprint = {2204.13714},
 primaryClass = {astro-ph.EP},
       adsurl = {https://ui.adsabs.harvard.edu/abs/2022ApJ...934...74M}
}

@ARTICLE{Pollack1996,
       author = {{Pollack}, James B. and {Hubickyj}, Olenka and {Bodenheimer}, Peter and {Lissauer}, Jack J. and {Podolak}, Morris and {Greenzweig}, Yuval},
        title = "{Formation of the Giant Planets by Concurrent Accretion of Solids and Gas}",
      journal = {\icarus},
         year = 1996,
        month = nov,
       volume = {124},
       number = {1},
        pages = {62-85},
          doi = {10.1006/icar.1996.0190},
       adsurl = {https://ui.adsabs.harvard.edu/abs/1996Icar..124...62P}
}

@ARTICLE{Batygin2023,
       author = {{Batygin}, Konstantin and {Morbidelli}, Alessandro},
        title = "{Formation of rocky super-earths from a narrow ring of planetesimals}",
      journal = {Nat. Astron.},
         year = 2023,
        month = mar,
       volume = {7},
        pages = {330-338},
          doi = {10.1038/s41550-022-01850-5},
archivePrefix = {arXiv},
       eprint = {2301.04680},
 primaryClass = {astro-ph.EP},
       adsurl = {https://ui.adsabs.harvard.edu/abs/2023NatAs...7..330B}
}

@ARTICLE{Ogihara2024,
       author = {{Ogihara}, Masahiro and {Morbidelli}, Alessandro and {Kunitomo}, Masanobu},
        title = "{Super-Earth Formation with Slow Migration from a Ring in an Evolving Peaked Disk Compatible with Terrestrial Planet Formation}",
      journal = {\apj},
         year = 2024,
        month = sep,
       volume = {972},
       number = {2},
          eid = {181},
        pages = {181},
          doi = {10.3847/1538-4357/ad65d5},
archivePrefix = {arXiv},
       eprint = {2407.15386},
 primaryClass = {astro-ph.EP},
       adsurl = {https://ui.adsabs.harvard.edu/abs/2024ApJ...972..181O}
}

@ARTICLE{Johansen2019,
       author = {{Johansen}, Anders and {Bitsch}, Bertram},
        title = "{Exploring the conditions for forming cold gas giants through planetesimal accretion}",
      journal = {\aap},
         year = 2019,
        month = nov,
       volume = {631},
          eid = {A70},
        pages = {A70},
          doi = {10.1051/0004-6361/201936351},
archivePrefix = {arXiv},
       eprint = {1909.10429},
 primaryClass = {astro-ph.EP},
       adsurl = {https://ui.adsabs.harvard.edu/abs/2019A&A...631A..70J}
}

@ARTICLE{Bitsch2023giants,
       author = {{Bitsch}, Bertram and {Izidoro}, Andre},
        title = "{Giants are bullies: How their growth influences systems of inner sub-Neptunes and super-Earths}",
      journal = {\aap},
         year = 2023,
        month = jun,
       volume = {674},
          eid = {A178},
        pages = {A178},
          doi = {10.1051/0004-6361/202245040},
archivePrefix = {arXiv},
       eprint = {2304.12758},
 primaryClass = {astro-ph.EP},
       adsurl = {https://ui.adsabs.harvard.edu/abs/2023A&A...674A.178B}
}

@ARTICLE{Bitsch2015,
       author = {{Bitsch}, Bertram and {Lambrechts}, Michiel and {Johansen}, Anders},
        title = "{The growth of planets by pebble accretion in evolving protoplanetary discs}",
      journal = {\aap},
         year = 2015,
        month = oct,
       volume = {582},
          eid = {A112},
        pages = {A112},
          doi = {10.1051/0004-6361/201526463},
archivePrefix = {arXiv},
       eprint = {1507.05209},
 primaryClass = {astro-ph.EP},
       adsurl = {https://ui.adsabs.harvard.edu/abs/2015A&A...582A.112B}
}

@ARTICLE{Levison2015,
       author = {{Levison}, Harold F. and {Kretke}, Katherine A. and {Duncan}, Martin J.},
        title = "{Growing the gas-giant planets by the gradual accumulation of pebbles}",
      journal = {\nat},
         year = 2015,
        month = aug,
       volume = {524},
       number = {7565},
        pages = {322-324},
          doi = {10.1038/nature14675},
archivePrefix = {arXiv},
       eprint = {1510.02094},
 primaryClass = {astro-ph.EP},
       adsurl = {https://ui.adsabs.harvard.edu/abs/2015Natur.524..322L}
}

@ARTICLE{Feinstein2025,
       author = {{Feinstein}, Adina D. and {Booth}, Richard A. and {Bergner}, Jennifer B. and {Lothringer}, Joshua D. and {Matthews}, Elisabeth C. and {Welbanks}, Luis and {Miguel}, Yamila and {Bitsch}, Bertram and {Eriksson}, Linn E.~J. and {Kirk}, James and {Pelletier}, Stefan and {Penzlin}, Anna B.~T. and {Piette}, Anjali A.~A. and {Piaulet-Ghorayeb}, Caroline and {Schwarz}, Kamber and {Turrini}, Diego and {Acu{\~n}a-Aguirre}, Lorena and {Ahrer}, Eva-Maria and {Barber}, Madyson G. and {Brande}, Jonathan and {Chakrabarty}, Aritra and {Crossfield}, Ian J.~M. and {Marleau}, Gabriel-Dominique and {Huang}, Helong and {Johansen}, Anders and {Kreidberg}, Laura and {Livingston}, John H. and {Luque}, Rafael and {Oreshenko}, Maria and {Pacetti}, Elenia and {Perotti}, Guilia and {Polman}, Jesse and {Prinoth}, Bibiana and {Semenov}, Dmitry A. and {Simon}, Jacob B. and {Teske}, Johanna and {Whiteford}, Niall},
        title = "{On Linking Planet Formation Models, Protoplanetary Disk Properties, and Mature Gas Giant Exoplanet Atmospheres}",
      journal = {arXiv e-prints},
         year = 2025,
        month = may,
          eid = {arXiv:2506.00669},
        pages = {arXiv:2506.00669},
          doi = {10.48550/arXiv.2506.00669},
archivePrefix = {arXiv},
       eprint = {2506.00669},
 primaryClass = {astro-ph.EP},
       adsurl = {https://ui.adsabs.harvard.edu/abs/2025arXiv250600669F}
}

@ARTICLE{Schneider2024,
       author = {{Schneider}, Aaron David and {Bitsch}, Bertram},
        title = "{chemcomp: Modeling the chemical composition of planets formed in protoplanetary disks}",
      journal = {arXiv e-prints},
         year = 2023,
        month = nov,
          eid = {arXiv:2401.15686},
        pages = {arXiv:2401.15686},
          doi = {10.48550/arXiv.2401.15686},
archivePrefix = {arXiv},
       eprint = {2401.15686},
 primaryClass = {astro-ph.IM},
       adsurl = {https://ui.adsabs.harvard.edu/abs/2024arXiv240115686S}
}

@ARTICLE{Pavlyuchenkov2007,
       author = {{Pavlyuchenkov}, Ya. and {Dullemond}, C.~P.},
        title = "{Dust crystallinity in protoplanetary disks: the effect of diffusion/viscosity ratio}",
      journal = {\aap},
         year = 2007,
        month = sep,
       volume = {471},
       number = {3},
        pages = {833-840},
          doi = {10.1051/0004-6361:20077317},
archivePrefix = {arXiv},
       eprint = {0706.2614},
 primaryClass = {astro-ph},
       adsurl = {https://ui.adsabs.harvard.edu/abs/2007A&A...471..833P}
}

@ARTICLE{Weidenschilling1977,
       author = {{Weidenschilling}, S.~J.},
        title = "{Aerodynamics of solid bodies in the solar nebula.}",
      journal = {\mnras},
         year = 1977,
        month = jul,
       volume = {180},
        pages = {57-70},
          doi = {10.1093/mnras/180.2.57},
       adsurl = {https://ui.adsabs.harvard.edu/abs/1977MNRAS.180...57W}
}

@ARTICLE{Brauer2008,
       author = {{Brauer}, F. and {Dullemond}, C.~P. and {Henning}, Th.},
        title = "{Coagulation, fragmentation and radial motion of solid particles in protoplanetary disks}",
      journal = {\aap},
         year = 2008,
        month = mar,
       volume = {480},
       number = {3},
        pages = {859-877},
          doi = {10.1051/0004-6361:20077759},
archivePrefix = {arXiv},
       eprint = {0711.2192},
 primaryClass = {astro-ph},
       adsurl = {https://ui.adsabs.harvard.edu/abs/2008A&A...480..859B}
}

@ARTICLE{Ndugu2024,
       author = {{Ndugu}, N. and {Bitsch}, B. and {Lienert}, J.~L.},
        title = "{How external photoevaporation changes the chemical composition of the inner disc}",
      journal = {\aap},
         year = 2024,
        month = nov,
       volume = {691},
          eid = {A32},
        pages = {A32},
          doi = {10.1051/0004-6361/202451633},
archivePrefix = {arXiv},
       eprint = {2409.07596},
 primaryClass = {astro-ph.EP},
       adsurl = {https://ui.adsabs.harvard.edu/abs/2024A&A...691A..32N}
}

@ARTICLE{Paardekooper2006,
       author = {{Paardekooper}, S. -J. and {Mellema}, G.},
        title = "{Dust flow in gas disks in the presence of embedded planets}",
      journal = {\aap},
         year = 2006,
        month = jul,
       volume = {453},
       number = {3},
        pages = {1129-1140},
          doi = {10.1051/0004-6361:20054449},
archivePrefix = {arXiv},
       eprint = {astro-ph/0603132},
 primaryClass = {astro-ph},
       adsurl = {https://ui.adsabs.harvard.edu/abs/2006A&A...453.1129P}
}

@ARTICLE{Paardekooper2014,
       author = {{Paardekooper}, S. -J.},
        title = "{Dynamical corotation torques on low-mass planets}",
      journal = {\mnras},
         year = 2014,
        month = nov,
       volume = {444},
       number = {3},
        pages = {2031-2042},
          doi = {10.1093/mnras/stu1542},
archivePrefix = {arXiv},
       eprint = {1409.0372},
 primaryClass = {astro-ph.EP},
       adsurl = {https://ui.adsabs.harvard.edu/abs/2014MNRAS.444.2031P}
}

@ARTICLE{BenitezLlambay2015,
       author = {{Ben{\'\i}tez-Llambay}, Pablo and {Masset}, Fr{\'e}d{\'e}ric and {Koenigsberger}, Gloria and {Szul{\'a}gyi}, Judit},
        title = "{Planet heating prevents inward migration of planetary cores}",
      journal = {\nat},
         year = 2015,
        month = apr,
       volume = {520},
       number = {7545},
        pages = {63-65},
          doi = {10.1038/nature14277},
archivePrefix = {arXiv},
       eprint = {1510.01778},
 primaryClass = {astro-ph.EP},
       adsurl = {https://ui.adsabs.harvard.edu/abs/2015Natur.520...63B}
}

@ARTICLE{Jimenez2017,
       author = {{Jim{\'e}nez}, Mar{\'\i}a Alejandra and {Masset}, Fr{\'e}d{\'e}ric S.},
        title = "{Improved torque formula for low- and intermediate-mass planetary migration}",
      journal = {\mnras},
         year = 2017,
        month = nov,
       volume = {471},
       number = {4},
        pages = {4917-4929},
          doi = {10.1093/mnras/stx1946},
archivePrefix = {arXiv},
       eprint = {1707.08988},
 primaryClass = {astro-ph.EP},
       adsurl = {https://ui.adsabs.harvard.edu/abs/2017MNRAS.471.4917J}
}

@ARTICLE{Crida2007,
       author = {{Crida}, A. and {Morbidelli}, A.},
        title = "{Cavity opening by a giant planet in a protoplanetary disc and effects on planetary migration}",
      journal = {\mnras},
         year = 2007,
        month = may,
       volume = {377},
       number = {3},
        pages = {1324-1336},
          doi = {10.1111/j.1365-2966.2007.11704.x},
archivePrefix = {arXiv},
       eprint = {astro-ph/0703151},
 primaryClass = {astro-ph},
       adsurl = {https://ui.adsabs.harvard.edu/abs/2007MNRAS.377.1324C}
}

@ARTICLE{BergezCasalou2020,
       author = {{Bergez-Casalou}, C. and {Bitsch}, B. and {Pierens}, A. and {Crida}, A. and {Raymond}, S.~N.},
        title = "{Influence of planetary gas accretion on the shape and depth of gaps in protoplanetary discs}",
      journal = {\aap},
         year = 2020,
        month = nov,
       volume = {643},
          eid = {A133},
        pages = {A133},
          doi = {10.1051/0004-6361/202038304},
archivePrefix = {arXiv},
       eprint = {2010.00485},
 primaryClass = {astro-ph.EP},
       adsurl = {https://ui.adsabs.harvard.edu/abs/2020A&A...643A.133B}
}

@INPROCEEDINGS{Baruetau2014,
       author = {{Baruteau}, C. and {Crida}, A. and {Paardekooper}, S.-J. and {Masset}, F. and {Guilet}, J. and {Bitsch}, B. and {Nelson}, R. and {Kley}, W. and {Papaloizou}, J.},
        title = "{Planet-Disk Interactions and Early Evolution of Planetary Systems}",
    booktitle = {Protostars and Planets VI},
         year = 2014,
       editor = {{Beuther}, Henrik and {Klessen}, Ralf S. and {Dullemond}, Cornelis P. and {Henning}, Thomas},
        month = jan,
        pages = {667-689},
          doi = {10.2458/azu_uapress_9780816531240-ch029},
archivePrefix = {arXiv},
       eprint = {1312.4293},
 primaryClass = {astro-ph.EP},
       adsurl = {https://ui.adsabs.harvard.edu/abs/2014prpl.conf..667B}
}

@ARTICLE{Lodders2003,
       author = {{Lodders}, Katharina},
        title = "{Solar System Abundances and Condensation Temperatures of the Elements}",
      journal = {\apj},
         year = 2003,
        month = jul,
       volume = {591},
       number = {2},
        pages = {1220-1247},
          doi = {10.1086/375492},
       adsurl = {https://ui.adsabs.harvard.edu/abs/2003ApJ...591.1220L}
}

@ARTICLE{Johansen2007,
       author = {{Johansen}, Anders and {Oishi}, Jeffrey S. and {Mac Low}, Mordecai-Mark and {Klahr}, Hubert and {Henning}, Thomas and {Youdin}, Andrew},
        title = "{Rapid planetesimal formation in turbulent circumstellar disks}",
      journal = {\nat},
         year = 2007,
        month = aug,
       volume = {448},
       number = {7157},
        pages = {1022-1025},
          doi = {10.1038/nature06086},
archivePrefix = {arXiv},
       eprint = {0708.3890},
 primaryClass = {astro-ph},
       adsurl = {https://ui.adsabs.harvard.edu/abs/2007Natur.448.1022J}
}

@ARTICLE{Youdin2007,
       author = {{Youdin}, A. and {Johansen}, A.},
        title = "{Protoplanetary Disk Turbulence Driven by the Streaming Instability: Linear Evolution and Numerical Methods}",
      journal = {\apj},
         year = 2007,
        month = jun,
       volume = {662},
       number = {1},
        pages = {613-626},
          doi = {10.1086/516729},
archivePrefix = {arXiv},
       eprint = {astro-ph/0702625},
 primaryClass = {astro-ph},
       adsurl = {https://ui.adsabs.harvard.edu/abs/2007ApJ...662..613Y}
}

@ARTICLE{Youdin2005,
       author = {{Youdin}, Andrew N. and {Goodman}, Jeremy},
        title = "{Streaming Instabilities in Protoplanetary Disks}",
      journal = {\apj},
         year = 2005,
        month = feb,
       volume = {620},
       number = {1},
        pages = {459-469},
          doi = {10.1086/426895},
archivePrefix = {arXiv},
       eprint = {astro-ph/0409263},
 primaryClass = {astro-ph},
       adsurl = {https://ui.adsabs.harvard.edu/abs/2005ApJ...620..459Y}
}

@INPROCEEDINGS{Drazkowska2023,
       author = {{Dr{\k{a}}{\.z}kowska}, J. and {Bitsch}, B. and {Lambrechts}, M. and {Mulders}, G.~D. and {Harsono}, D. and {Vazan}, A. and {Liu}, B. and {Ormel}, C.~W. and {Kretke}, K. and {Morbidelli}, A.},
        title = "{Planet Formation Theory in the Era of ALMA and Kepler: from Pebbles to Exoplanets}",
    booktitle = {Protostars and Planets VII},
         year = 2023,
       editor = {{Inutsuka}, S. and {Aikawa}, Y. and {Muto}, T. and {Tomida}, K. and {Tamura}, M.},
       series = {Astronomical Society of the Pacific Conference Series},
       volume = {534},
        month = jul,
        pages = {717},
          doi = {10.48550/arXiv.2203.09759},
archivePrefix = {arXiv},
       eprint = {2203.09759},
 primaryClass = {astro-ph.EP},
       adsurl = {https://ui.adsabs.harvard.edu/abs/2023ASPC..534..717D}
}

@ARTICLE{Johansen2019mig,
       author = {{Johansen}, Anders and {Ida}, Shigeru and {Brasser}, Ramon},
        title = "{How planetary growth outperforms migration}",
      journal = {\aap},
         year = 2019,
        month = feb,
       volume = {622},
          eid = {A202},
        pages = {A202},
          doi = {10.1051/0004-6361/201834071},
archivePrefix = {arXiv},
       eprint = {1811.00523},
 primaryClass = {astro-ph.EP},
       adsurl = {https://ui.adsabs.harvard.edu/abs/2019A&A...622A.202J}
}

@ARTICLE{Oberg2021,
       author = {{{\"O}berg}, Karin I. and {Guzm{\'a}n}, Viviana V. and {Walsh}, Catherine and {Aikawa}, Yuri and {Bergin}, Edwin A. and {Law}, Charles J. and {Loomis}, Ryan A. and {Alarc{\'o}n}, Felipe and {Andrews}, Sean M. and {Bae}, Jaehan and {Bergner}, Jennifer B. and {Boehler}, Yann and {Booth}, Alice S. and {Bosman}, Arthur D. and {Calahan}, Jenny K. and {Cataldi}, Gianni and {Cleeves}, L. Ilsedore and {Czekala}, Ian and {Furuya}, Kenji and {Huang}, Jane and {Ilee}, John D. and {Kurtovic}, Nicolas T. and {Le Gal}, Romane and {Liu}, Yao and {Long}, Feng and {M{\'e}nard}, Fran{\c{c}}ois and {Nomura}, Hideko and {P{\'e}rez}, Laura M. and {Qi}, Chunhua and {Schwarz}, Kamber R. and {Sierra}, Anibal and {Teague}, Richard and {Tsukagoshi}, Takashi and {Yamato}, Yoshihide and {van't Hoff}, Merel L.~R. and {Waggoner}, Abygail R. and {Wilner}, David J. and {Zhang}, Ke},
        title = "{Molecules with ALMA at Planet-forming Scales (MAPS). I. Program Overview and Highlights}",
      journal = {\apjs},
         year = 2021,
        month = nov,
       volume = {257},
       number = {1},
          eid = {1},
        pages = {1},
          doi = {10.3847/1538-4365/ac1432},
archivePrefix = {arXiv},
       eprint = {2109.06268},
 primaryClass = {astro-ph.EP},
       adsurl = {https://ui.adsabs.harvard.edu/abs/2021ApJS..257....1O}
}

@INPROCEEDINGS{Paardekooper2023,
       author = {{Paardekooper}, S. and {Dong}, R. and {Duffell}, P. and {Fung}, J. and {Masset}, F.~S. and {Ogilvie}, G. and {Tanaka}, H.},
        title = "{Planet-Disk Interactions and Orbital Evolution}",
    booktitle = {Protostars and Planets VII},
         year = 2023,
       editor = {{Inutsuka}, S. and {Aikawa}, Y. and {Muto}, T. and {Tomida}, K. and {Tamura}, M.},
       series = {Astronomical Society of the Pacific Conference Series},
       volume = {534},
        month = jul,
        pages = {685},
          doi = {10.48550/arXiv.2203.09595},
archivePrefix = {arXiv},
       eprint = {2203.09595},
 primaryClass = {astro-ph.EP},
       adsurl = {https://ui.adsabs.harvard.edu/abs/2023ASPC..534..685P}
}

@ARTICLE{EvansSoma2025,
       author = {{Evans-Soma}, Thomas M. and {Sing}, David K. and {Barstow}, Joanna K. and {Piette}, Anjali A.~A. and {Taylor}, Jake and {Lothringer}, Joshua D. and {Reggiani}, Henrique and {Goyal}, Jayesh M. and {Ahrer}, Eva-Maria and {Mayne}, Nathan J. and {Rustamkulov}, Zafar and {Kataria}, Tiffany and {Christie}, Duncan A. and {Gapp}, Cyril and {Dong}, Jiayin and {Foreman-Mackey}, Daniel and {Hattori}, Soichiro and {Marley}, Mark S.},
        title = "{SiO and a super-stellar C/O ratio in the atmosphere of the giant exoplanet WASP-121 b}",
      journal = {Nat. Astron.},
         year = 2025,
        month = jun,
       volume = {9},
        pages = {845-861},
          doi = {10.1038/s41550-025-02513-x},
archivePrefix = {arXiv},
       eprint = {2506.01771},
 primaryClass = {astro-ph.EP},
       adsurl = {https://ui.adsabs.harvard.edu/abs/2025NatAs...9..845E}
}

@ARTICLE{Bardet2025,
       author = {{Bardet}, Deborah and {Changeat}, Quentin and {Venot}, Olivia and {Panek}, Emilie},
        title = "{Re-analysis of 10 Hot-Jupiter Atmospheres with disequilibrium chemistry retrieval}",
      journal = {arXiv e-prints},
         year = 2025,
        month = jun,
          eid = {arXiv:2506.12806},
        pages = {arXiv:2506.12806},
          doi = {10.48550/arXiv.2506.12806},
archivePrefix = {arXiv},
       eprint = {2506.12806},
 primaryClass = {astro-ph.EP},
       adsurl = {https://ui.adsabs.harvard.edu/abs/2025arXiv250612806B}
}

@ARTICLE{Hoch2023,
       author = {{Hoch}, Kielan K.~W. and {Konopacky}, Quinn M. and {Theissen}, Christopher A. and {Ruffio}, Jean-Baptiste and {Barman}, Travis S. and {Rickman}, Emily L. and {Perrin}, Marshall D. and {Macintosh}, Bruce and {Marois}, Christian},
        title = "{Assessing the C/O Ratio Formation Diagnostic: A Potential Trend with Companion Mass}",
      journal = {\aj},
         year = 2023,
        month = sep,
       volume = {166},
       number = {3},
          eid = {85},
        pages = {85},
          doi = {10.3847/1538-3881/ace442},
archivePrefix = {arXiv},
       eprint = {2212.04557},
 primaryClass = {astro-ph.EP},
       adsurl = {https://ui.adsabs.harvard.edu/abs/2023AJ....166...85H}
}

@ARTICLE{Changeat2022,
       author = {{Changeat}, Q. and {Edwards}, B. and {Al-Refaie}, A.~F. and {Tsiaras}, A. and {Skinner}, J.~W. and {Cho}, J.~Y.~K. and {Yip}, K.~H. and {Anisman}, L. and {Ikoma}, M. and {Bieger}, M.~F. and {Venot}, O. and {Shibata}, S. and {Waldmann}, I.~P. and {Tinetti}, G.},
        title = "{Five Key Exoplanet Questions Answered via the Analysis of 25 Hot-Jupiter Atmospheres in Eclipse}",
      journal = {\apjs},
         year = 2022,
        month = may,
       volume = {260},
       number = {1},
          eid = {3},
        pages = {3},
          doi = {10.3847/1538-4365/ac5cc2},
archivePrefix = {arXiv},
       eprint = {2204.11729},
 primaryClass = {astro-ph.EP},
       adsurl = {https://ui.adsabs.harvard.edu/abs/2022ApJS..260....3C}
}

@ARTICLE{Pacetti2025,
       author = {{Pacetti}, E. and {Schisano}, E. and {Turrini}, D. and {Dullemond}, C.~P. and {Molinari}, S. and {Walsh}, C. and {Fonte}, S. and {Lebreuilly}, U. and {Klessen}, R.~S. and {Hennebelle}, P. and {Ivanovski}, S.~L. and {Politi}, R. and {Polychroni}, D. and {Simonetti}, P. and {Testi}, L.},
        title = "{Planet formation in chemically diverse and evolving discs: I. Composition of planetary building blocks}",
      journal = {\aap},
         year = 2025,
        month = sep,
       volume = {701},
          eid = {A194},
        pages = {A194},
          doi = {10.1051/0004-6361/202554012},
archivePrefix = {arXiv},
       eprint = {2506.17399},
 primaryClass = {astro-ph.EP},
       adsurl = {https://ui.adsabs.harvard.edu/abs/2025A&A...701A.194P}
}

@ARTICLE{Cridland2019,
       author = {{Cridland}, Alex J. and {van Dishoeck}, Ewine F. and {Alessi}, Matthew and {Pudritz}, Ralph E.},
        title = "{Connecting planet formation and astrochemistry. A main sequence for C/O in hot exoplanetary atmospheres}",
      journal = {\aap},
         year = 2019,
        month = dec,
       volume = {632},
          eid = {A63},
        pages = {A63},
          doi = {10.1051/0004-6361/201936105},
archivePrefix = {arXiv},
       eprint = {1910.13171},
 primaryClass = {astro-ph.EP},
       adsurl = {https://ui.adsabs.harvard.edu/abs/2019A&A...632A..63C}
}

@ARTICLE{Tabone2023,
       author = {{Tabone}, B. and {Bettoni}, G. and {van Dishoeck}, E.~F. and {Arabhavi}, A.~M. and {Grant}, S. and {Gasman}, D. and {Henning}, Th. and {Kamp}, I. and {G{\"u}del}, M. and {Lagage}, P.~O. and {Ray}, T. and {Vandenbussche}, B. and {Abergel}, A. and {Absil}, O. and {Argyriou}, I. and {Barrado}, D. and {Boccaletti}, A. and {Bouwman}, J. and {Caratti o Garatti}, A. and {Geers}, V. and {Glauser}, A.~M. and {Justannont}, K. and {Lahuis}, F. and {Mueller}, M. and {Nehm{\'e}}, C. and {Olofsson}, G. and {Pantin}, E. and {Scheithauer}, S. and {Waelkens}, C. and {Waters}, L.~B.~F.~M. and {Black}, J.~H. and {Christiaens}, V. and {Guadarrama}, R. and {Morales-Calder{\'o}n}, M. and {Jang}, H. and {Kanwar}, J. and {Pawellek}, N. and {Perotti}, G. and {Perrin}, A. and {Rodgers-Lee}, D. and {Samland}, M. and {Schreiber}, J. and {Schwarz}, K. and {Colina}, L. and {{\"O}stlin}, G. and {Wright}, G.},
        title = "{A rich hydrocarbon chemistry and high C to O ratio in the inner disk around a very low-mass star}",
      journal = {Nat. Astron.},
         year = 2023,
        month = jul,
       volume = {7},
        pages = {805-814},
          doi = {10.1038/s41550-023-01965-3},
archivePrefix = {arXiv},
       eprint = {2304.05954},
 primaryClass = {astro-ph.EP},
       adsurl = {https://ui.adsabs.harvard.edu/abs/2023NatAs...7..805T}
}

@ARTICLE{vantHoff2020,
       author = {{van 't Hoff}, Merel L.~R. and {Bergin}, Edwin A. and {J{\o}rgensen}, Jes K. and {Blake}, Geoffrey A.},
        title = "{Carbon-grain Sublimation: A New Top-down Component of Protostellar Chemistry}",
      journal = {\apjl},
         year = 2020,
        month = jul,
       volume = {897},
       number = {2},
          eid = {L38},
        pages = {L38},
          doi = {10.3847/2041-8213/ab9f97},
       adsurl = {https://ui.adsabs.harvard.edu/abs/2020ApJ...897L..38V}
}

@ARTICLE{Polychroni2025,
       author = {{Polychroni}, D. and {Turrini}, D. and {Ivanovski}, S. and {Marzari}, F. and {Testi}, L. and {Politi}, R. and {Sozzetti}, A. and {Trigo-Rodriguez}, J.~M. and {Desidera}, S. and {Drozdovskaya}, M.~N. and {Fonte}, S. and {Molinari}, S. and {Naponiello}, L. and {Pacetti}, E. and {Schisano}, E. and {Simonetti}, P. and {Zusi}, M.},
        title = "{HD 163296 and its giant planets: Creation of exo-comets, interstellar objects and transport of volatile material}",
      journal = {\aap},
         year = 2025,
        month = may,
       volume = {697},
          eid = {A158},
        pages = {A158},
          doi = {10.1051/0004-6361/202453097},
archivePrefix = {arXiv},
       eprint = {2503.04669},
 primaryClass = {astro-ph.EP},
       adsurl = {https://ui.adsabs.harvard.edu/abs/2025A&A...697A.158P}
}

\begin{appendix}

\section{Influence of other disc parameters}\label{appendix_other_parameters}

\begin{figure*}
    \centering
    \includegraphics[width=0.9\textwidth]{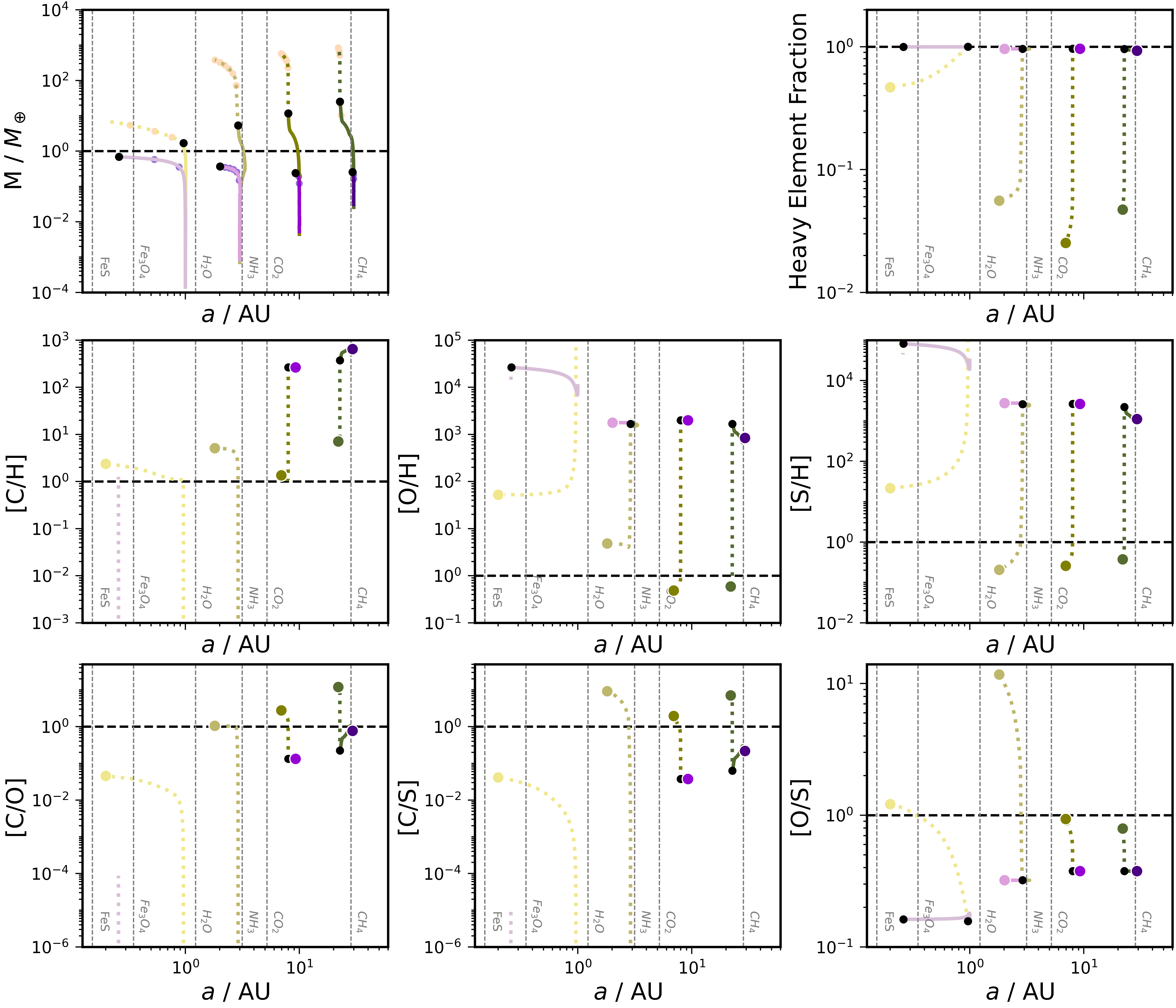}
    \caption{Same as Fig.~\ref{fig:growth_ratios_ufrag}, but now the purple colour represents planets growing in discs with $M_{\rm disc}=0.01M_\odot$ rather than $M_{\rm disc}=0.1M_\odot$ (green colour).}    
    \label{growth_ratios_M0}
\end{figure*}

\subsection{Lower disc mass} 
\label{app:discmass}

As shown in the top left panel of Fig. \ref{growth_ratios_M0}, with a reduced disc mass, the expected outcome is a lower planetary mass due to the limited material available for formation. Therefore, the planets may not reach the gas accretion phase, resulting in a thinner or absent H-He envelope. Consequently, the elemental ratios within the atmosphere are dominated by the evaporated material during core build-up, resulting in very extreme [S/H], [O/H], and [C/H] ratios (middle row of Fig.~\ref{growth_ratios_M0}). These can either be very high or very low, depending on whether hydrogen can be accreted during the core build-up (e.g. via CH$_4$, H$_2$O, NH$_3$ or H$_2$S). The formed planets in these two different sets of simulations are very different (gas giants with extended H/He envelope or core-dominated planets with a small heavy element envelope). Consequently, the atmospheric abundances differ by a few orders of magnitude, which is also reflected by the [C/O], [C/S], and [O/S] ratios (bottom row in Fig.~\ref{growth_ratios_M0}) and their total heavy element fractions (top right in Fig.~\ref{growth_ratios_M0}).

\subsection{Later starting time of embryo} 
\label{app:startingtime}

\begin{figure*}[htbp]
    \centering
    \includegraphics[width=0.9\textwidth]{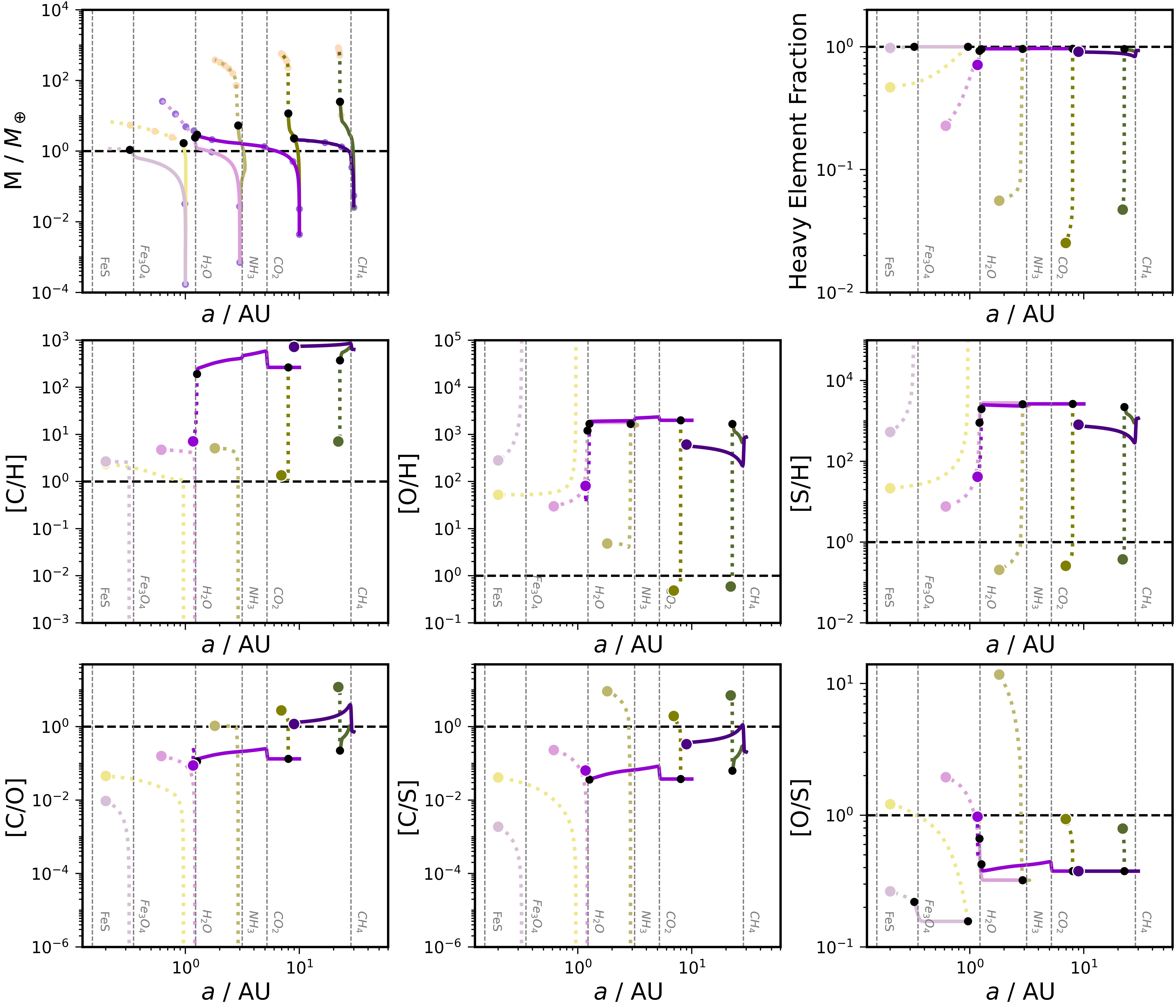}
    \caption{Same as Fig.~\ref{fig:growth_ratios_ufrag}, but now the purple colour represents planets starting to grow at $t_0=0.9$ Myr rather than $t_0=0.1$ Myr (green colour).}    
    \label{growth_ratios_t0}
\end{figure*}

The later starting time of the planetary embryos results in reduced growth ( top left panel in Fig.~\ref{growth_ratios_t0}), as the majority of the pebbles have drifted inwards and thus reduce the accretion rates \citep{Savvidou_2023}. Consequently, the planetary growth tracks are very different. While the outermost planet, starting at 30 AU, does not grow beyond a few Earth masses, it migrates significantly in type-I migration, even in the low viscosity environment. The planet forming at 10 AU experiences a similar fate, that it migrates interior to 1 AU due to efficient type-I migration and only reaching around 5 Earth masses at the end of the simulation. Only the planet starting at 3 AU starts to accrete gas, but stays at a sub-Saturn level.

As our simulations do not include a temperature evolution, the pebble isolation mass remains constant in time. Consequently, if the planets reach the pebble isolation mass, they will be the same mass as planets forming earlier in the disc. However, due to the slower growth, the planets forming later migrate more and consequently cross more evaporation fronts, leading to a difference in their elemental abundances during the core growth phase (middle and bottom row of Fig~\ref{growth_ratios_t0}) as well as their heavy element mass fraction (top right in Fig.~\ref{growth_ratios_t0}). This shows again the importance of the migration pathway the planet takes, rather than when it starts to grow.

\subsection{Smaller disc radius} 
\label{app:discradius}

\begin{figure*}[htbp]
    \centering
    \includegraphics[width=0.9\textwidth]{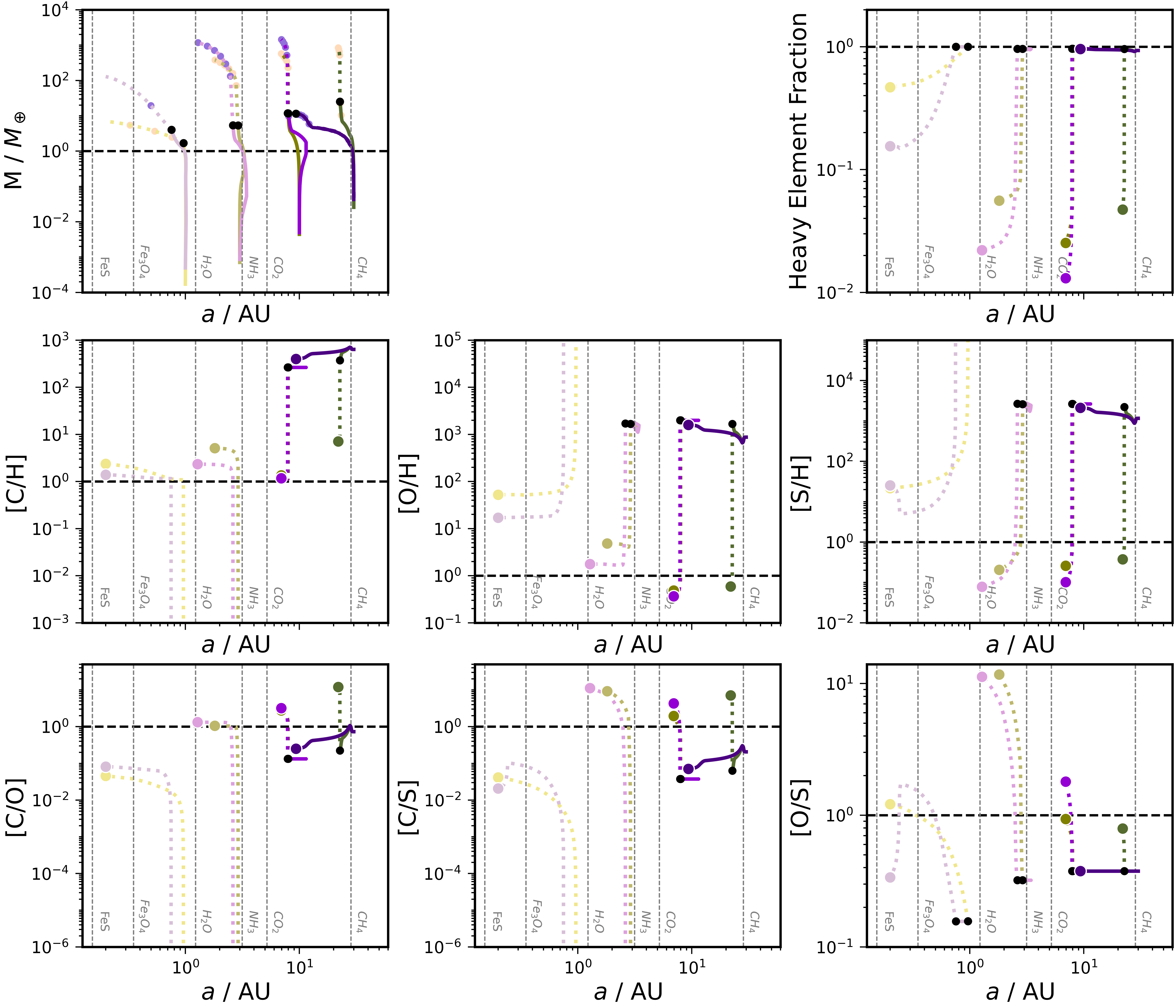}
    \caption{Same as Fig.~\ref{fig:growth_ratios_ufrag}, but now the purple colour represents planets growing in discs with $R_0=50$AU rather than $R_0=200$ AU (green colour).}        
    \label{growth_ratios_R0}
\end{figure*}

Fig.~\ref{growth_ratios_R0} shows the evolution of planets growing in discs with smaller radii. As the disc radius is smaller, its gas surface density is higher (as the overall disc mass is the same). Consequently, the gas accretion rates are higher, as they are limited by the disc's accretion rate ($\propto \nu \Sigma$), and can result in higher planetary masses. This is visible for the planets forming at 3 and 10 AU, where the planets forming in the smaller disc grow more massive (top left panel in Fig.~\ref{growth_ratios_R0}) by roughly a factor of $\sim 2$.

On the other hand, the migration pathway for the planets starting at 3 and 10 AU is similar and the planets stay within the same evaporation fronts, resulting in atmospheric abundances and ratios that agree within a factor of $\sim 2$. This relatively large difference is caused by the fact that the more massive planets accrete relatively more hydrogen compared to volatiles.

The growth pattern for the planets forming at 1 and 30 AU show significant differences. The core of the planet starting at 30 AU does not grow efficiently, because the small disc size will result in a very short-lived pebble flux, which hinders the growth of planets (e.g. \citealt{Levison2015}). Consequently, the planet remains at a few Earth masses - too low to start to accrete a gaseous envelope efficiently - and migrates inwards to 10 AU. Thus, the corresponding atmospheric abundances differ by a few orders of magnitude. We note here that the duration of the pebble flux is longer for the planets growing in the inner disc regions, because pebbles starting from the outer disc have to travel a longer distance (e.g. \citealt{lambrechts2014forming}). The planet growing at 1 AU, on the other hand, accretes a larger planetary envelope compared to its counterpart in the disc with a larger radius. This is caused by the fact that the planet can accrete more material initially from its horseshoe region (due to the enhanced disc density) and during the runaway accretion phase. Consequently, the atmospheric abundances of these planets differ by a factor of a few (middle and bottom row in Fig.~\ref{growth_ratios_R0}. The heavy element mass fraction shows similar differences (top right in Fig.~\ref{growth_ratios_R0}.

\subsection{Larger dust-to-gas ratio} 
\label{app:dustratio}

\begin{figure*}[htbp]
    \centering
    \includegraphics[width=0.9\textwidth]{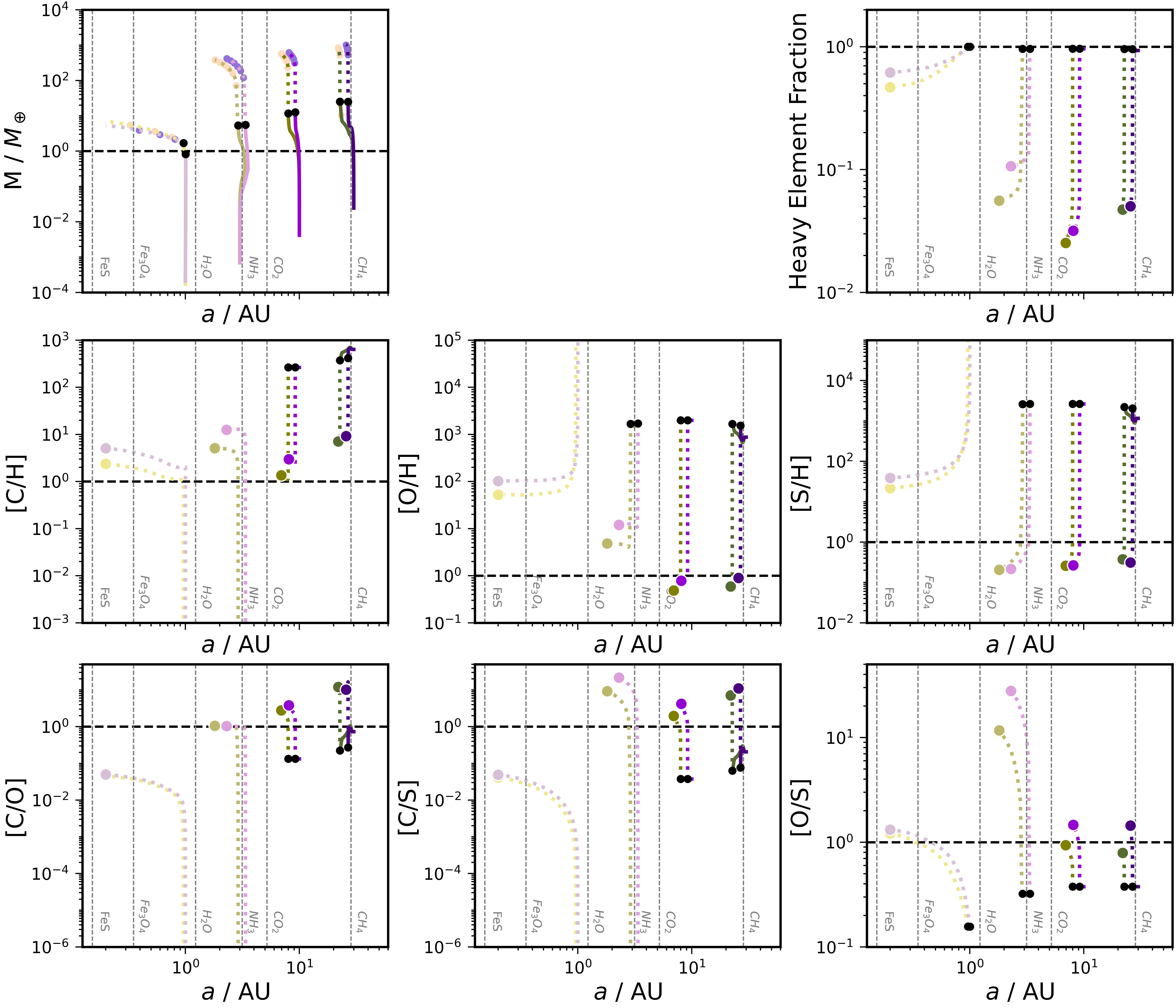}
    \caption{Same as Fig.~\ref{fig:growth_ratios_ufrag}, but now the purple colour represents planets growing in discs with a larger dust-to-gas ratio of $f_{\rm DG}=0.03$ rather than $f_{\rm DG}=0.015$ (green colour).}        
    \label{growth_ratios_dtg}
\end{figure*}

Fig.~\ref{growth_ratios_dtg} shows the evolution of the planet in discs with different dust-to-gas ratios. As the larger dust-to-gas ratio allows for a more efficient core build-up phase, it consequently also leads to a reduced type-I migration phase, as the planets are basically outgrowing migration \citep{Johansen2019mig}. 

As the planetary migration pathway is very similar - due to the low viscosity - the planets remain within the same evaporation fronts during their growth for the different simulations (top left in Fig.~\ref{growth_ratios_dtg}). However, the [C/H] and [O/H] fractions are enhanced by a factor of two compared to our standard simulations (middle row in Fig.~\ref{growth_ratios_dtg}). The larger dust-to-gas ratio implies more inward drifting pebbles that evaporate and enrich the disc with more volatiles that are consequently accreted by gas accreting planets. Consequently, the planets forming in discs with a higher dust-to-gas ratio result in planets that are more enriched in carbon and oxygen compared to our standard simulations. In our case, the dust-to-gas ratio is larger by a factor of two and thus the planets are enriched by a factor of about two compared to our standard simulations. For sulphur, the situation is different, because sulphur is accreted as a solid into the planet, where the amount of material does not matter, but its ratio to hydrogen does. However, the ratio of sulphur to the other elements remains constant, meaning that the composition ratios of the solids in the disc are unaffected by the larger dust-to-gas ratio. Consequently, the planetary atmospheres have basically identical sulphur abundances, except for the planets forming interior to the water ice line, where a small fraction of sulphur (carried in $H_2S$) evaporates and can be accreted with the gas. As oxygen and carbon are the main carriers of the heavy elements in the disc, the total heavy element content of the planets also shows an enlargement up to a factor of two compared to our standard simulations (top right in Fig.~\ref{growth_ratios_dtg}). 

As carbon and oxygen are enhanced both by a factor of two, the [C/O] ratio remains the same at higher dust-to-gas ratio, but the [C/S] and [O/S] ratio differ by a factor of two (bottom row in Fig.~\ref{growth_ratios_dtg}). We, however, want to point out that if only one elemental ratio were changed (e.g. more oxygen), the corresponding ratios containing oxygen (e.g. [C/O]) will change as well. This points to the fact that the detailed stellar abundances are necessary if one wants to constrain the planet formation pathway via atmospheric abundance measurements.

\section{C/H, O/H and S/H of planetary atmospheres}
\label{ap:ratios}

\begin{figure*}[htbp]
    \centering
    \includegraphics[width=0.9\textwidth]{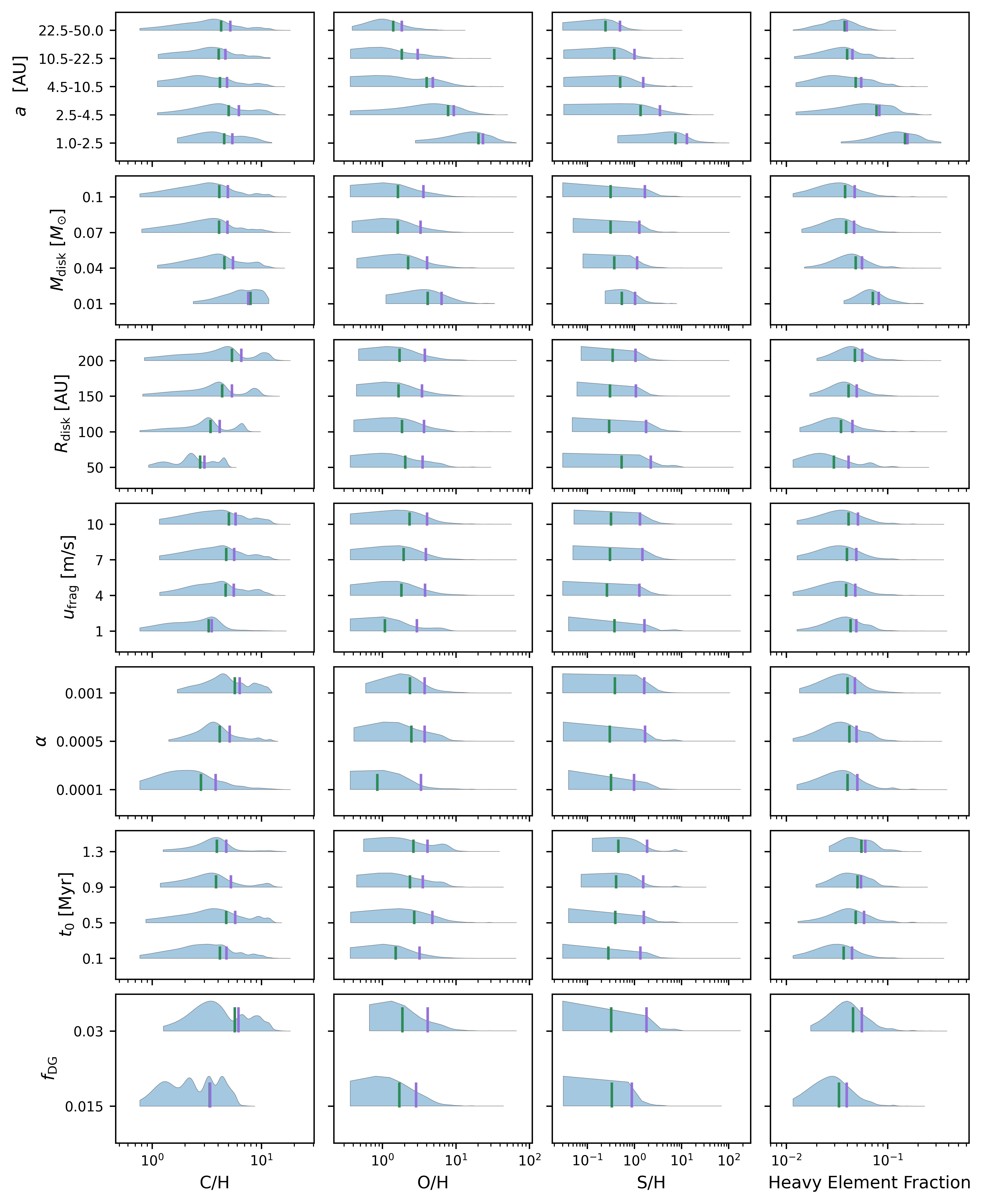}
    \caption{Atmospheric C/H, O/H, S/H (all normalised to solar values) and heavy element fraction (left to right) of planets above 100 Earth masses in our sample as a function of the individually probed disc and planet parameters (top to bottom). The green and blue vertical lines mark the median and mean of the distribution, respectively.}
    \label{fig:violin_H}
\end{figure*}

We show in Fig.~\ref{fig:violin_H} the distributions of the atmospheric C/H, O/H, and S/H as well as the heavy element fraction (left to right) of planets above 100 Earth masses in our sample for the different probed parameters (top to bottom). Oxygen, carbon and sulphur probe different accretion processes of our simulations. While oxygen is mostly volatile, carbon is entirely volatile, with a maximum evaporation temperature of 70K for CO$_2$, whereas sulphur is 90\% refractory in our model (set by FeS). Consequently, sulphur can only be accreted in small amounts during the core build-up as well as through the gas interior of 150K (H$_2$S evaporation). At colder temperatures - so further outside of a few AU - sulphur can only be accreted into the planetary core during core formation. Once the core has reached its pebble isolation mass, it blocks all sulphur, which is stored in FeS and H$_2$S pebbles at these temperatures. Consequently, planets forming in the very outer regions of the disc will have a low sulphur abundance. We remind that we attribute 10\% of the initially accreted core mass to the atmosphere, explaining why the formed planets show an atmospheric sulphur abundance.

In contrast, oxygen and carbon can be accreted with the gas phase at all orbital distances interior to the CO evaporation front. This corresponds to all planets in our sample (see Fig.~\ref{fig:mass_vs_distance_ratios}), as the CO evaporation front is exterior to 50 AU for all disc models.

The individual elemental abundances of the planetary atmospheres depend, therefore, differently on the initial position (top row in Fig.~\ref{fig:violin_H}). The S/H abundances are mostly sub-stellar for planets forming in the outer disc regions (exterior to 2.5 AU), as sulphur can only be accreted during the core build-up. Once planets migrate interior to the H$_2$S evaporation front, their sulphur content can become super-stellar. Only if the planets migrate into the very inner regions, where FeS evaporates and releases sulphur into the gas phase, can very high S/H values be achieved for the planetary atmospheres.

A similar behaviour is seen for the oxygen abundances of the planetary atmospheres. Planets forming and staying in the outer regions have a relatively low oxygen abundance that can be sub-stellar, whereas planets that migrate into the inner disc regions and interior to the CO$_2$ and H$_2$O evaporation front have super-stellar O/H. In particular, planets forming within 3 AU always show a super-stellar O/H content.

In contrast, the C/H ratio always seems super-stellar for nearly all planets in our sample. This is caused by the fact that almost all planets, even if they form in the very outer regions, migrate across the CH$_4$ evaporation front, which increases the carbon fraction significantly. The exceptions are planets forming exterior to the CH$_4$ evaporation front in low viscosity environments, where migration is slow (Fig.~\ref{fig:discparam}). The mean and median distribution of the C/H ratios are otherwise very similar for all formation locations, which is caused by the fact that carbon species evaporate all exterior to a few AU. We note that the inclusion of refractory carbon grains and their consequent destruction in the inner disc region  \citep{Houge2025} could change the overall C/H distribution of planetary atmospheres. However, considering that our simulations show significant trends with the atmospheric abundances already, a change in where different materials evaporate would be recovered by the simulations.

On the other hand, the disc properties themselves have an influence on the C/H, O/H, and S/H, as in the case of the C/O ratio. We will discuss now the influence of the different parameters on the C/O, C/H, and O/H in more detail before we discuss the heavy element fraction of the planets.

\subsection{Disc mass}

C/H, O/H, and S/H show similar trends with the disc mass (2nd row in Fig.~\ref{fig:violin_H}). While the mean and averages of the distributions remain similar for disc masses above 4\% $M_\odot$, planets forming in lower mass discs feature larger atmospheric abundances for oxygen and carbon. This is caused by the fact that low mass discs (1\% $M_\odot$) only allow the formation of giant planets for our simulations with enhanced dust-to-gas ratios, explaining the overall larger minimum abundances in carbon, oxygen, and sulphur. In addition, these planets mostly form in the inner disc regions (Fig.~\ref{fig:discparam}), where most of the heavy elements can be accreted with the gas. In contrast, in the higher mass discs, planets can also form in the outer disc regions, where not all volatiles have evaporated, resulting in a lower limit of the atmospheric abundances for the planets formed in these discs.

We note that the sulphur abundance for planets forming exterior to the H$_2$S line is entirely accreted during the core build-up, when solids can still be accreted. As planets can form further out in more massive discs, the number of these outer planets increases. At the same time, the sulphur fraction inside the planets decreases at larger distances to the star, as other species condense out (e.g. CO$_2$, CH$_4$, CO), reducing the overall fraction of solid sulphur that is accreted for cores of the same mass.

\subsection{Disc radius}

The C/H ratio shows a clear trend with disc radius (3rd row from top in Fig.~\ref{fig:violin_H}). The smallest discs produce planets with a lower median and mean C/H content compared to the larger discs. This is caused by the simple fact that planets can not form efficiently in the outer disc regions (exterior to 30ish AU) if the disc only has a critical radius of 50 AU \citep{Savvidou_2023}. Therefore, fewer planets form in the outer regions of the disc, where carbon is enhanced due to the evaporation of CH$_4$. This effect also explains the 2\textsuperscript{nd} peak of the C/H distribution at around $\sim5-10$ times stellar: it is caused by planets forming in the outer regions close to the CH$_4$ evaporation front, resulting in the formation of planets that are enriched in C/H.

The oxygen and sulphur abundances, on the other hand, do not show a strong trend with disc radius. This is caused by the fact that essentially all sulphur is accreted during the solid phase, so similarly for all planets. Oxygen is already accreted with CO, which evaporates in the outer disc regions and is thus accessible for all form planets (to some level), resulting in a similar O/H distribution for discs with different radii.

\subsection{Grain fragmentation velocity}

The grain fragmentation velocity sets the sizes of the pebbles in our model. We probe velocities ranging from 1 to 10 m/s (4th row from top in Fig.~\ref{fig:violin_H}), which results in a size difference of a factor of 100 for these two extreme values \citep{birnstiel2012simple}. We see that the C/H and O/H distributions of the formed planets don't feature large super-stellar values for a fragmentation velocity of 1m/s. In this case, the enrichment of the gas phase with vapour is lower, because the pebbles drift inwards slower and therefore the reduced velocity difference between pebbles and gas reduces the enrichment (e.g. \citealt{mah2023close}). For fragmentation velocities of 4m/s and above, the distributions of O/H and C/H are similar, as the difference in pebble velocity is just a factor of a few. Again, for sulphur, the differences are negligible, as sulphur is mostly accreted into the planets by solids during the core build-up, which is then distributed into the envelope (see above). While the grain fragmentation velocity changes the grain sizes, it does not change the elemental ratios within the pebbles. Thus, even if the solid accretion rate might differ, the composition of the solid material that is accreted during core build-up stays the same, resulting in the same S/H abundance ratios for all grain fragmentation velocities.

\subsection{Viscosity}

The viscosity has two main effects in our model: i) it sets how fast giant planets can migrate, and ii) how fast the vapour is transported inwards. We see that planets formed in environments with lower viscosity have, on average, lower C/H and O/H ratios, while the S/H ratio is not strongly affected (5th row from top in Fig.~\ref{fig:violin_H}). The lower C/H ratios can be explained by the fact that planets forming exterior to the CH$_4$ line do not cross it in low viscosity environments due to their slow migration (see Fig.~\ref{fig:discparam}). Consequently, fewer planets will have a large enrichment with carbon, compared to planets forming in discs with higher viscosities, where migration is more efficient. 

A similar argument can be made for O/H. Here, the slower migration caused by low viscosity prevents planets from forming exterior to 3-5 AU to cross the water ice line. This reduces the number of planets that will be enriched in oxygen compared to the case with higher viscosities. 

The sulphur abundance, on the other hand, shows no clear trend with viscosity, as the mean and median values are very similar in all cases. This is caused again by the properties of sulphur, which is mostly refractory and is thus accreted mostly during core build-up, essentially unaffected by viscosity.

\subsection{Planetary starting time}

The planetary starting time of the embryos seems to have only minimal influence on the composition of planets (2nd panel from bottom in Fig.~\ref{fig:violin_H}). Earlier formation times allow the formation of more carbon-rich planets, because the enhancement of the disc with vapour is largest at early times and planets can form in the very outer disc regions, where the C/H is largest. On the other hand, the O/H distribution of the planets forming at 1.3 Myr shows an enhancement of the planets that are enriched by a factor of a few. This is caused by the fact that at late starting times, only planets in the inner disc regions, where the oxygen abundance is highest, can grow to become gas giants \citep{Savvidou_2023}. Consequently, there are fewer giant planets forming in the outer regions, reducing the number of planets that have lower oxygen abundances in their atmosphere, giving rise to the bump at O/H$\sim$5-10. Sulphur follows the same trend, that planets forming later seem to have larger sulphur abundances. This is again caused by the fact that the late-forming planets form in the inner disc and thus closer to the FeS evaporation front, allowing a more efficient accretion of sulphur vapour into their atmospheres.

\subsection{Dust-to-gas ratio}

The dust-to-gas ratio influences the atmospheric composition of the planets (bottom row in Fig.~\ref{fig:violin_H}), because as more heavy elements are available in the disc, the planet has more opportunities to accrete them. In particular, the abundances of carbon and oxygen in the atmospheres of the planets are enhanced, as these materials can be accreted efficiently in gaseous form in all disc regions.

On the other hand, the sulphur abundances cover a similar range. This is caused by the fact that sulphur is mostly accreted in solid form during the core build-up of the planet. However, the relative mixture of the different heavy elements remains the same at higher dust-to-gas ratios, because the higher dust-to-gas ratio only sets the total amount of material, but not its composition. Planets migrating into the very inner disc regions can also accrete sulphur via the gas from the evaporated FeS. This then causes a tail towards higher S/H ratios (around 2-10 times stellar), which is not visible at low dust-to-gas ratios.

\subsection{Heavy element fraction}

The right column of Fig.~\ref{fig:violin_H} shows the heavy element fraction of the planets above 100 Earth masses in our sample. The overall shape of the heavy element fraction distributions across all disc parameters is very similar to the distribution of the planetary C/H and O/H. This is caused by the fact that the large, heavy element contents of the planets are generally dominated by the accretion of vapour-enriched material, traced by the planetary C/H and O/H ratios, rather than by solid accretion, which is set by the core mass.

The initial starting position of the planet seems to have the strongest influence on the heavy element fraction of the giant planets. In particular, planets forming in the inner disc (interior to 2.5 AU) have a larger heavy element fraction (with a mean of $\sim 15\%$), but also a larger total heavy element mass (see Fig.~\ref{fig:mass_vs_distance_ratios}). The heavy element fraction of the giant planets is dominated by the accretion of gas enriched in vapour. As this enrichment is larger in the inner disc regions \citep{bitsch2023enriching}, the planets forming in the inner disc regions have a larger heavy element fraction compared to the planets forming in the outer disc regions, where the mean heavy element fraction reduces to a few percentage.

In particular, as expected, an overall larger disc dust-to-gas ratio results in more enriched planets (bottom right in Fig.~\ref{fig:violin_H}). The mean heavy element fraction seems to increase with the starting time of the planetary embryos. This is a bit counter-intuitive, as the heavy element fraction in the gas phase decreases with disc age \citep{bitsch2023enriching}. However, at later times ($t_0=1.3$ Myr), planets only form in the inner disc regions \citep{Savvidou_2023}, where the vapour enhancement is larger compared to the outer disc, resulting in a larger mean/media heavy element content. However, the maximal heavy element content that the giant planets in our simulations can reach is larger for early formation due to the larger vapour enrichment in the inner disc.

Lower disc masses seem to produce giant planets with larger heavy element mass fractions. This is caused by two effects: (i) the giant planets in these discs are generally smaller (see Fig.~\ref{fig:discparam}) meaning that the core has a larger effect on the heavy element mass fraction and (ii) the giant planets form predominately in the inner disc regions, where the enrichment with heavy element is larger compared to the outer disc regions. On the other hand, the disc's viscosity, the grain fragmentation velocity, and the disc radius only seem to have minimal influence on the heavy element fraction of the planets.

\section{Influence of the disc parameters}
\label{app:discparameters}

\begin{figure*}[htbp]
    \centering \includegraphics[width=1.0\textwidth]{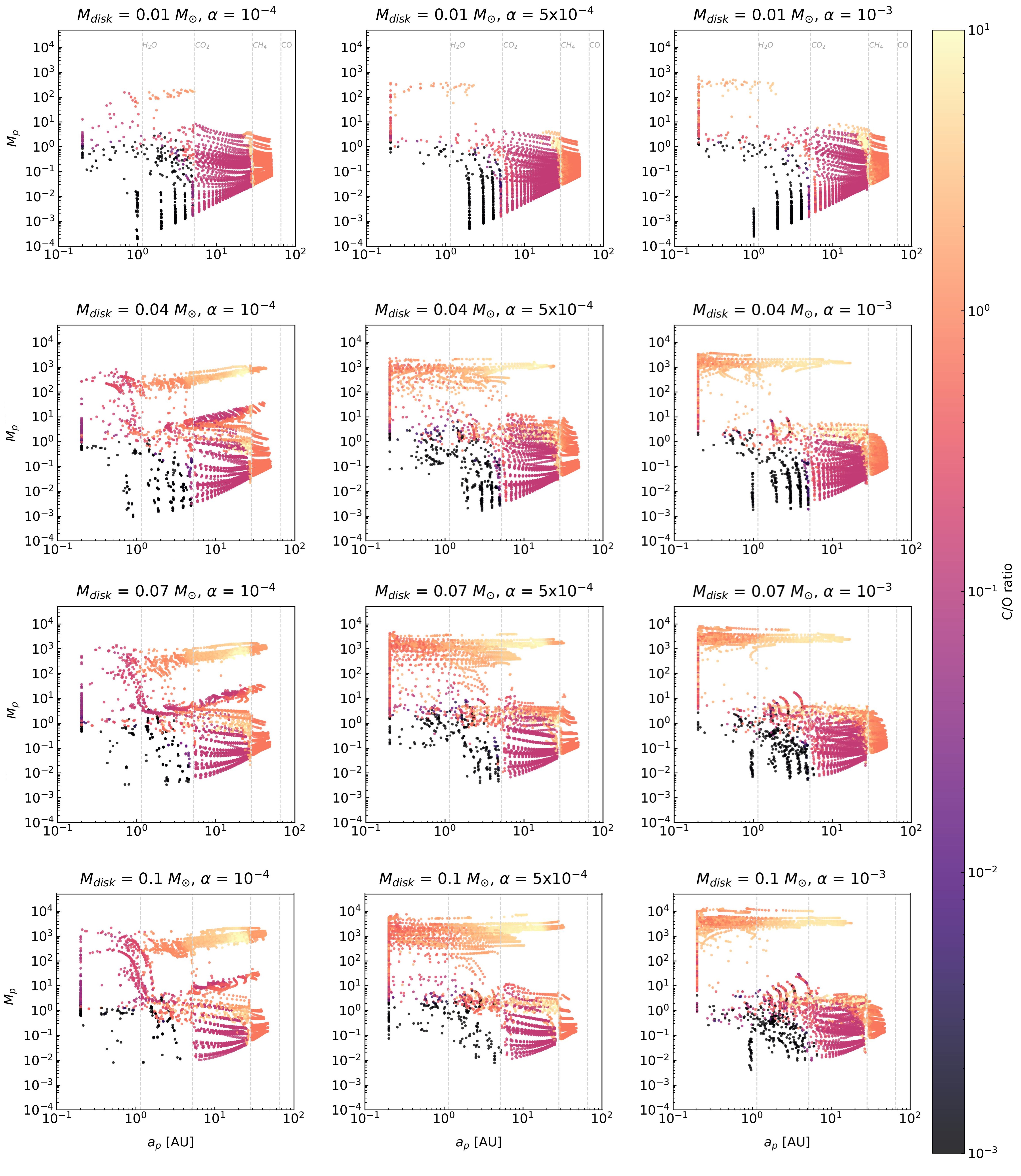} 
    \caption{Final planetary masses as a function of their final positions, colour coded by the atmospheric C/O ratio of the planets normalised to the stellar value. The different columns correspond to simulations with different viscosities, while the different rows show our probed initial disc masses. The vertical lines correspond to the position of the evaporation fronts of the major volatile carriers (H$_2$O, CO$_2$, CH$_4$ and CO).}
    \label{fig:discparam}
\end{figure*}

We investigated how the grain fragmentation velocity and the disc's viscosity influence the formation and composition of giant planets (Fig.~\ref{fig:growth_ratios_ufrag} and Fig.~\ref{fig:growth_ratios_alpha}). While we found only small differences in the atmospheric abundances (of the order of 10-20\%) for varying grain fragmentation velocities, we found larger differences (a factor of a few) in the case of varying disc viscosities. This is caused by the fact that planets forming in discs with larger viscosities migrate more efficiently and thus cross more evaporation fronts, allowing the planets to accrete materials with different compositions. We now show in Fig.~\ref{fig:discparam} the final planetary masses as a function of the final planetary positions, colour-coded by the atmospheric C/O ratio for our whole sample, split by different disc masses and viscosities. This allows us to identify some of the previous trends in a clearer way.

As we found before, a very low disc mass ($\sim 1\% M_\odot$) does not allow the efficient formation of giant planets \citep{Savvidou_2023}, independently of the disc's viscosity (top row in Fig.~\ref{fig:discparam}), because the disc does not harbour enough solid materials to allow the efficient growth of embryos to the pebble isolation mass. If the disc mass is large enough to allow the formation of planets reaching the pebble isolation mass, a higher disc viscosity will allow the formation of more massive planets. This is caused by the fact that the planetary accretion rate is limited by the disc's accretion rate, which is proportional to the disc's viscosity.

As discussed above, a higher disc's viscosity results in more efficient inward migration of growing giant planets, resulting in changes in the planets' composition. Fig.~\ref{fig:discparam} shows this effect: planets forming in discs with higher viscosity migrate inwards further (right-hand panels in Fig.~\ref{fig:discparam}) compared to planets forming in discs with lower viscosity (left-hand panels in Fig.~\ref{fig:discparam}). This effect is visualized by the fact that the even though the planets start at the same position (every AU up to 50 AU), they end up closer to the host star (e.g. interior to the CH$_4$ evaporation front) in discs with high viscosity compared to planets forming in discs with low viscosity, where a clear planet population exists exterior to the CH$_4$ evaporation front with an atmospheric C/O ratio of unity.

However, when comparing the atmospheric C/O ratio of giant planets in discs with different disc masses but the same viscosities, a clear trend emerges: The atmospheric C/O ratio of the giant planets seems not to be affected by the disc mass to a large extent. It is instead determined by their position relative to the different evaporation fronts. Even in the case of high viscosity, where the planets migrate larger distances and cross multiple evaporation fronts, the individual disc mass only plays a minor role for the atmospheric C/O ratio, as long as one compares planets that end up between the same evaporation fronts and at the same mass.

\end{appendix}

\end{document}